\documentclass[12pt]{article}

\usepackage{amssymb,amsfonts,amsmath}
\usepackage{graphicx} 
\usepackage{indentfirst}

\usepackage{bbm}

\parskip=\medskipamount

\newcommand {\cA}{{\cal A}}
\newcommand {\cB}{{\cal B}}
\newcommand {\cC}{{\cal C}}
\newcommand {\cD}{{\cal D}}
\newcommand {\cE}{{\cal E}}

\newcommand {\cL}{{\cal L}}
\newcommand {\cM}{{\cal M}}
\newcommand {\cN}{{\cal N}}

\newcommand {\cR}{{\cal R}}
\newcommand {\cS}{{\cal S}}

\newcommand{\bH}{{\bf H}}

\def\a{\alpha}
\def \bi{\bibitem}

\def\b{\beta}

\def\d{\delta}
\def\e{\epsilon}

\def\g{\gamma}
\def\G{\Gamma}

\def\j{\psi}

\def\l{\lambda}

\def\p{\pi}

\def\r{\rho}
\def\s{\sigma}

\def\z{\zeta}
\def\D{\Delta}
\def\F{\Phi}
\def\J{\Psi}
\def\L{\Lambda}
\def\O{\Omega}

\def\S{\Sigma}
\def\U{\Upsilon}

\def\tr{{\rm tr}}
\newcommand{\ad}{{\dot{\alpha}}}                           %new
\newcommand{\pa}{\partial}                           %new
\newcommand{\hf}{\frac12}
\newcommand{\vf}{\varphi}
\newcommand{\sect}[1]{\setcounter{equation}{0}\section{#1}}

\newcommand{\be}{\begin{equation}}
\newcommand{\ee}{\end{equation}}
\newcommand{\bea}{\begin{eqnarray}}
\newcommand{\eea}{\end{eqnarray}}
\newcommand{\non}{\nonumber}

\def\dt#1{{\buildrel {\hbox{\LARGE .}} \over {#1}}}    % dot-over for sp/sb

\newcommand{\bm}[1]{\mbox{\boldmath$#1$}}

\def\double #1{#1{\hbox{\kern-2pt $#1$}}}

\begin{document}

\begin{titlepage}

\begin{flushright}
HIP-2006-54/TH\\
UUITP-20/06\\
hep-th/0612174\\
December, 2006\\
\end{flushright}
\vspace{5mm}

\begin{center}
{\Large \bf  Hyperk\"ahler sigma models on cotangent bundles} \\
{\Large \bf of Hermitian symmetric spaces using}  \\
{\Large \bf projective superspace}
\end{center}

\begin{center}

{\large  
Masato Arai${},^{a,}$\footnote{masato.arai@helsinki.fi}
Sergei M. Kuzenko${}^{b,}$\footnote{{kuzenko@cyllene.uwa.edu.au}}
and Ulf Lindstr\"om${}^{c,}$\footnote{ulf.lindstrom@teorfys.uu.se} 
} \\
\vspace{5mm}

\footnotesize{
${}^{a}${\it High Energy Physics Division, Department of Physical Sciences\\
University of Helsinki and Helsinki Institute of Physics\\
P.O. Box 64, FIN-00014, Finland
}}  
\\
\vspace{2mm}

\footnotesize{
${}^{b}${\it School of Physics M013, The University of Western Australia\\
35 Stirling Highway, Crawley W.A. 6009, Australia}}  
\\
\vspace{2mm}

\footnotesize{
${}^{c}${\it Department of Theoretical Physics,
Uppsala University \\ 
Box 803, SE-751 08 Uppsala, Sweden\\
 and\\
 Helsinki Institute of Physics\\
P.O. Box 64, FIN-00014, Finland
}}
\\

\vspace{2mm}

\end{center}
\vspace{5mm}

\begin{abstract}
\baselineskip=14pt
\noindent
K\"ahler manifolds have a natural hyperk\"ahler structure associated 
 with (part of) their cotangent bundles.
Using projective superspace, we construct four-dimensional ${\cal N}=2$ 
 models on the tangent bundles of some classical 
Hermitian symmetric spaces (specifically, the four regular series of irreducible 
compact symmetric K\"ahler manifolds, and their non-compact versions).
A further dualization yields the K\"ahler potential for the
 hyperk\"ahler metric on the cotangent bundle.
\end{abstract}
\vspace{1cm}

\vfill
\end{titlepage}

\newpage
\setcounter{page}{1}
\renewcommand{\thefootnote}{\arabic{footnote}}
\setcounter{footnote}{0}

\tableofcontents{}
\vspace{1cm}
\bigskip\hrule

\sect{Introduction}

Supersymmetry in sigma models is closely related to the  geometry of target space \cite{Zumino}. 
In particular,
$\cN =2$ models in 
four space-time dimensions
require the target space geometry to be hyperk\"ahler
\cite{Alvarez-Gaume:1981hm}. Consequently, constructions of new $\cN =2$ sigma
models lead to new hyperk\"ahler metrics, a fact which has been extensively pursued
in the Legendre transform and hyperk\"ahler quotient constructions 
\cite{Lindstrom:1983rt, Hitchin:1986ea}.  

To fully utilize the relation to geometry, manifest $\cN =2$ formulations are needed. 
Projective superspace 
\cite{Karlhede:1984vr, Hitchin:1986ea}
provides this, and has led to the discovery of a number of new multiplets 
that can be used to construct new hyperk\"ahler metrics (see  e.g. \cite{LR1}).  
In \cite{LR1} a generalized Legendre transform was devised,
that produces hyperk\"ahler metrics.\footnote{The term ``projective superspace''
was coined in   \cite{LR2}.}

Among the projective supermultiplets, perhaps the most interesting one is 
the so-called polar multiplet \cite{LR1} which can be used to describe a charged 
U(1) hypermultiplet coupled to a vector multiplet \cite{LR2},
and therefore
is analogous to
the $\cN=1$ chiral superfield. 
The polar multiplet\footnote{The terminology 
``polar'' and ``(ant)arctic'' multiplets was introduced in  
 \cite{G-RRWLvU}.} 
is described 
by an arctic superfield $\U (\z)$ and 
its complex conjugate composed with the antipodal map
$\bar{\z}\rightarrow -1/\z$, 
the antarctic superfield $\breve{\U} (\z)$.
It is required to possess certain holomorphy properties
 on a punctured two-plane parametrized by 
the complex variable $\z$ 
(the latter may be interpreted as a projective coordinate 
on $\mathbb{C}P^1$).
When realized in ordinary $\cN=1$ superspace,
$\U (\z)$ and  $\breve{\U} (\z) $ 
are generated by an infinite set of ordinary superfields:
\be
 \U (\z) = \sum_{n=0}^{\infty}  \, \U_n \z^n = 
\F + \S \,\z+ O(\z^2) ~,\qquad
\breve{\U} (\z) = \sum_{n=0}^{\infty}  \, {\bar
\U}_n
 (-\z)^{-n}~.
\label{exp}
\ee
Here $\F$ is chiral, $\S$  complex linear, 
\be
{\bar D}_{\dt{\a}} \F =0~, \qquad \qquad {\bar D}^2 \S = 0 ~,
\label{chiral+linear}
\ee
and the remaining component superfields are unconstrained complex 
superfields.  Using the polar multiplet, 
one can construct 
a family of 4D $\cN=2$ off-shell 
supersymmetric nonlinear sigma-models that are described in 
 $\cN=1$ superspace by the action
\bea
S[\U, \breve{\U}]  =  
\frac{1}{2\pi {\rm i}} \, \oint \frac{{\rm d}\z}{\z} \,  
 \int {\rm d}^8 z \, 
K \big( \U (\z), \breve{\U} (\z), \z  \big) ~,
\label{genact}
\eea
with the integration contour around the origin in $\mathbb C$.

The unconstrained superfields   $\U_2, \U_3, \dots$, and their conjugates,
appear in the action (\ref{genact})  without derivatives, 
and therefore 
they are purely auxiliary degrees of freedom. 
Their role is to ensure a linearly realized $\cN=2$ supersymmetry.  
In order to describe the theory only in terms of the physical superfields
$\U_0 =\F$ and $\U_1 =\S$, one has to eliminate all the auxiliary 
superfields using their equations of motion. 
The problem of elimination of the auxiliary superfields
is actually nontrivial, since one has to solve an infinite number 
of nonlinear equations. So far it has been solved
perturbatively only for a broad subclass of the models (\ref{genact})
 studied in \cite{K,GK1,GK2}, and also exact solutions have
been found in special cases \cite{GK1,GK2,AN,K3}.
Such a family is obtained
by restricting $K \big( \U, \breve{\U} ,\z \big) \to 
K \big( \U, \breve{\U}  \big)$ in  (\ref{genact}).
Then, the corresponding action can be viewed as a minimal  $\cN=2$ 
extension of the general four-dimensional $\cN=1$ supersymmetric nonlinear 
sigma model \cite{Zumino}, with $K(\F, \bar \F)$ the K\"ahler 
potential of a K\"ahler manifold $\cM$, 
and the physical superfields $(\F, \S)$ parametrizing 
the tangent bundle $T\cM$ of the K\"ahler manifold \cite{K}.
Upon elimination of the auxiliary superfields, 
the complex linear tangent variables $\S$ can be dualized 
 (as a final step of  the generalized Legendre transform
\cite{LR1}) into chiral one-forms, such that  the target space for the model obtained 
turns out to be (an open domain of the zero section of)
the cotangent bundle $T^*\cM$ of the K\"ahler manifold \cite{GK1,GK2}.

The perturbative procedure for the elimination of the auxiliary superfields, 
which was originally described in \cite{GK1}, 
has recently been refined by one of us (SMK), see e.g. \cite{KL}.
The scheme obtained
is reminiscent of the mathematical techniques 
used to prove  the theorem  \cite{cotangent} that, 
for a K\"ahler manifold $\cM$, a  canonical hyperk\"ahler structure 
 exists, in general, on an open neighborhood of the zero section 
of the cotangent bundle $T^*\cM$.

As outlined in \cite{GK1} and further elaborated in \cite{K3}, 
for Hermitian symmetric spaces 
the auxiliary fields may be eliminated exactly.
In the  present paper we make systematic use of this fact.  
The method entails finding a particular solution to 
the auxiliary field equations at the origin of $\cM$ 
(in a normal coordinate system\footnote{For such coordinates, 
the term ``K\"ahler normal coordinates'' was suggested in 
\cite{HN2}.}  introduced in \cite{A-GG,HKLR})
and then relying on the existence of holomorphic isometries 
and other special properties to extend the solution to an arbitrary  point. 
A key ingredient in this procedure is a
coset representative  of a convenient form.

Let us end the introduction with a brief comment on related work.
The first hyperk\"ahler manifolds  that are cotangent bundles of 
 certain complex Grassmannians  (generalizations of the Calabi
 manifolds), were presented in \cite{Lindstrom:1983rt}. 
Lately, massive versions of these and related models have been 
 discussed in \cite{Arai:2002xa, Arai:2003es, Arai:2003tc, Arai:2003my}
 using ${\cal N}=1$ superspace and $\cN=2$ harmonic superspace
 techniques \cite{Galperin:1984av}.
In addition to these specific examples, some structural results 
for massive $\cN=2$ sigma models have been obtained in 
$\cN=1$ superspace \cite{BX} and projective superspace \cite{K4}.

The presentation of the paper is organized as follows. 
Section 2 describes the type of $\cN = 2$ supersymmetric sigma models 
we are interested in and their background in projective superspace.  
In section 3 we construct the  coset representatives 
needed for the four types of compact Hermitian symmetric spaces. 
Section 4 contains the construction of the sigma models 
for a Grassmann manifold  and is  followed in sections 5 
and 6 by the construction for the remaining three symmetric spaces. 
In section 7 we repeat the 
discussions in section 3, but now for the non-compact versions of the
spaces.  The corresponding sigma
 models occupy sections 8, 9 and 10. 
In two appendices we present derivations of
 results needed in the general text. 

The results obtained in the present paper admit a natural extension
to 5D \cite{KL} and 6D \cite{GL,GPT} projective superspace
formulations.

\sect{Dynamical setup}\label{2}
This section sets the stage for our constructions of 
supersymmetric sigma models on symmetric spaces. 
In particular we introduce the relevant $\cN=2$ extensions of  $\cN=1$ sigma models 
and their origin in projective superspace.\\

Projective superspace is a superspace enlarged with a 
$\mathbb{C}P^1$ at  each point. The coordinates thus 
include an additional projective coordinate $\zeta$ 
on this space. The larger space allows for integrations over invariant subspaces, much 
like  the chiral  integrals in ordinary superspace, and hence for manifest  $\cN=2$ actions.
The integration measure contains a contour integral over a closed curve in the complex 
$\zeta$-plane which picks out the residue from  the Lagrangian.
Projective superspace was introduced in 
\cite{Karlhede:1984vr,Hitchin:1986ea}, 
and has been continuously developed over the years. 
For a recent mathematically oriented description which also 
elaborates the close relation to twistor space, see \cite{LR51}.\\

In this paper, we are interested in a family of 4D $\cN=2$ off-shell 
supersymmetric nonlinear sigma-models that are described in 
ordinary $\cN=1$ 
superspace by the action
\bea
S[\U, \breve{\U}]  =  
\frac{1}{2\pi {\rm i}} \, \oint \frac{{\rm d}\z}{\z} \,  
 \int {\rm d}^8 z \, 
K \big( \U^I (\z), \breve{\U}^{\bar{I}} (\z)  \big) ~.
\label{nact} 
\eea
These dynamical systems 
present themselves 
 a subclass of the more general 
family of 4D $\cN=2$ off-shell 
supersymmetric nonlinear models \cite{LR1}
in projective superspace, given in eq. (\ref{genact}).
What is special about the model (\ref{nact}) is its interesting geometric
properties 
\cite{K,GK1,GK2}. It occurs as a minimal $\cN=2$ extension of the
general four-dimensional $\cN=1$ supersymmetric nonlinear sigma model \cite{Zumino}
\be
S[\F, \bar \F] =  \int {\rm d}^8 z \, K(\Phi^{I},
 {\bar \Phi}{}^{\bar{J}})  ~,
\label{nact4}
\ee
with $K$ the  K\"ahler potential of a K\"ahler manifold $\cM$.

The extended supersymmetric  sigma model  (\ref{nact}) 
inherits  all the geometric features of
its $\cN=1$ predecessor (\ref{nact4}). 
The K\"ahler invariance of the latter,
\be
K(\F, \bar \F) \quad \longrightarrow \quad K(\F, \bar \F)~ +~
\L(\F)+  {\bar \L} (\bar \F) 
\label{kahl}
\ee
turns into 
\be
K(\U, \breve{\U})  \quad \longrightarrow \quad K(\U, \breve{\U}) ~+~
\L(\U) \,+\, {\bar \L} (\breve{\U} ) 
\label{kahl2}
\ee
for the model (\ref{nact}). 
A holomorphic reparametrization of the K\"ahler manifold,
\be
 \F^I  \quad  \longrightarrow   \quad f^I \big( \F \big) ~,
\ee
has the following
counterpart
\be
\U^I (\z) \quad  \longrightarrow  \quad f^I \big (\U(\z) \big)
\label{kahl3}
\ee
in the $\cN=2$ case. Therefore, the physical
superfields of the 
${\cal N}=2$ theory
\be
 \U^I (\z)\Big|_{\z=0} ~=~ \F^I ~,\qquad  \quad \frac{ {\rm d} \U^I (\z) 
}{ {\rm d} \z} \Big|_{\z=0} ~=~ \S^I ~,
\label{kahl4} 
\ee
should be regarded, respectively, as  coordinates of a point in the K\" ahler
manifold and a tangent vector at  the same point. 
Thus the variables $(\F^I, \S^J)$ parametrize the tangent 
bundle $T\cM$ of the K\"ahler manifold $\cM$ \cite{K}. 

To describe the theory in terms of 
the physical superfields $\F$ and $\S$ only, 
all the auxiliary 
superfields have to be eliminated  with the aid of the 
corresponding algebraic equations of motion
\bea
\oint \frac{{\rm d} \z}{\z} \,\z^n \, \frac{\pa K(\U, \breve{\U} 
) }{\pa \U^I} ~ = ~ \oint \frac{{\rm d} \z}{\z} \,\z^{-n} \, \frac{\pa 
K(\U, \breve{\U} ) } {\pa \breve{\U}^{\bar I} } ~ = ~
0 ~, \qquad n \geq 2 ~ .               
\label{asfem}
\eea
Let $\U_*(\z) \equiv \U_*( \z; \F, {\bar \F}, \S, \bar \S )$ 
denote a unique solution subject to the initial conditions
\bea
\U_* (0)  = \F ~,\qquad  \quad \dt{\U}_* (0) 
 = \S ~.
\label{geo3} 
\eea

For a general K\"ahler manifold $\cM$, 
the auxiliary superfields $\U_2, \U_3, \dots$, and their
conjugates,  can be eliminated  only perturbatively. 
Their elimination  can be carried out
using the ansatz \cite{KL}
\bea
\U^I_n = 
\sum_{p=0}^{\infty} 
G^I{}_{J_1 \dots J_{n+p} \, \bar{L}_1 \dots  \bar{L}_p} (\F, {\bar \F})\,
\S^{J_1} \dots \S^{J_{n+p}} \,
{\bar \S}^{ {\bar L}_1 } \dots {\bar \S}^{ {\bar L}_p }~, 
\qquad n\geq 2~.
\eea
Upon elimination of the auxiliary superfields,
the action 
(\ref{nact}) takes the form \cite{GK1,GK2}
\bea
S_{{\rm tb}}[\F, \bar \F, \S, \bar \S]  
&=& \int {\rm d}^8 z \, \Big\{\,
K \big( \F, \bar{\F} \big) - g_{I \bar{J}} \big( \F, \bar{\F} 
\big) \S^I {\bar \S}^{\bar{J}} 
\non\\
&&\qquad +
\sum_{p=2}^{\infty} \cR_{I_1 \cdots I_p {\bar J}_1 \cdots {\bar 
J}_p }  \big( \F, \bar{\F} \big) \S^{I_1} \dots \S^{I_p} 
{\bar \S}^{ {\bar J}_1 } \dots {\bar \S}^{ {\bar J}_p }~\Big\}~, \label{act-tab}
\eea
where the tensors $\cR_{I_1 \cdots I_p {\bar J}_1 \cdots {\bar 
J}_p }$ are functions of the Riemann curvature $R_{I {\bar 
J} K {\bar L}} \big( \F, \bar{\F} \big) $ and its covariant 
derivatives.  Each term in the action contains equal powers
of $\S$ and $\bar \S$, since the original model (\ref{nact}) 
is invariant under rigid U(1)  transformations \cite{GK1}
\be
\U(\zeta) ~~ \mapsto ~~ \U({\rm e}^{{\rm i} \a} \zeta) 
\quad \Longleftrightarrow \quad 
\U_n(z) ~~ \mapsto ~~ {\rm e}^{{\rm i} n \a} \U_n(z) ~.
\label{rfiber}
\ee

${}$The process of eliminating the auxiliary fields from the action (\ref{nact})  and subsequently performing 
a Legendre transform with respect to linear fields is called 
a generalized Legendre transform
\cite{LR1}. For the theory with action  $S_{{\rm tb}}[\F, \bar \F, \S, \bar \S]  $, 
this gives a dual formulation involving only chiral superfields 
and their  conjugates as the dynamical variables. 
Consider the first-order action 
\bea
 S_{{\rm tb}}[\F, \bar \F, \S, \bar \S] 
+ \int {\rm d}^8 z \,
\Big\{ 
 \j_I \,\S^I + {\bar \j}_{\bar I} {\bar \S}^{\bar I} 
\Big\} ~,
\eea
where the tangent vector $\S^I$ is now  complex unconstrained, 
while the one-form $\j_I$ is chiral, ${\bar D}_{\dt \a} \j_I =0$.
Upon elimination of $\G$ and $\bar \G$, with the aid of their equations of 
motion, the action turns into 
\be
S_{{\rm cb}}[\F, \bar \F, \j, \bar \j]
=  \int {\rm d}^8 z \, H(\F, \bar \F, \j, \bar \j )~. 
\ee
Its target space is (an open domain of the zero section of)
 the cotangent bundle $T^*\cM$ of the K\"ahler manifold $\cM$, 
and $H(\F, \bar \F, \j, \bar \j )$ the corresponding hyperk\"ahler 
potential \cite{GK1,GK2}.

${}$For Hermitian symmetric spaces, the auxiliary superfields
can in principle be eliminated exactly, as outlined in \cite{GK1}. 
Here we present a more elaborated procedure following 
mainly \cite{K3}.

Given a K\"ahler manifold $\cM$, and an arbitrary  point $p_0 \in \cM$,   
one can construct a K\"ahler normal coordinate system 
with the origin at $p_0$  \cite{A-GG,HKLR}
(see also \cite{HN2} for a more recent discussion). 
Such a system  
is characterized by the conditions imposed at $p_0$:
\bea
K_{I_1 \dots I_n \, {\bar J}}  &=& K_{I \,{\bar J}_1 \dots {\bar J}_n} =0~,
\qquad n>1~,
\non \\
K_{I_1 \dots I_n }  &=& K_{{\bar J}_1 \dots {\bar J}_n} =0~, \non \\
K_{I {\bar J}}&=& \d_{I {\bar J}}~.
\eea
The specific feature of 
Hermitian symmetric  spaces is that, in addition, 
more conditions hold:
\bea
K_{I_1\dots I_m  \,{\bar J}_1 \dots {\bar J}_n} =0~,
\qquad m \neq n~.
\eea
These conditions imply that the K\"ahler potential is invariant 
under arbitrary U(1) phase transformations, 
$K({\rm e}^{ {\rm i} \a} \, \F ,  {\rm e}^{- {\rm i} \a}\, {\bar \F}) = 
K(\F , \bar \F )$, and therefore
$
K(\F, \bar \F) = F (\F \,\bar \F),
$
with $F (\F \,\bar \F)$ a real analytic function.

In accordance with  \cite{GK1,GK2},
for any Hermitian symmetric space $\cM$  
one can find  
the solution 
 $\U_*(\z) \equiv \U_*( \z; \F, {\bar \F}, \S, \bar \S )$
to the equations of motion (\ref{asfem}) 
in a closed form.
Let $\U_0(\z)$  denote  the value of $\U_* (\z)$ at the origin 
of the K\"ahler normal coordinate system, 
$\U_0(\z) =   \U_*( \z; \F =0, {\bar \F} =0 , \S_0, {\bar \S}_0 )$, 
with $\S_0$ a tangent vector at the origin. 
It is easy to check that 
\bea
\U_0 (\z) = \S_0 \, \z~, \qquad  
\breve{\U}_0 (\z) =-  \frac{\bar{\S}_0}{\z}
\label{u-0}
\eea
solve the equations (\ref{asfem}) at $\F=0$ and 
respect the initial conditions.
A  next step is to distribute this solution to any point $\F$
of the manifold $\cM$, that is to  make use 
of $ \U_*( \z; \F =0, {\bar \F} =0 , \S_0, {\bar \S}_0 )$ 
in order to obtain $\U_*( \z; \F, {\bar \F}, \S, \bar \S )$.

Let $G$ be the isometry group 
of the Hermitian symmetric space $\cM$.
It acts transitively on $\cM$ by holomorphic transformations. 
Without loss of generality, we can always choose the open domain  $U$, 
on which the K\"ahler normal coordinate system is defined, 
to be simply connected. Then, we can construct 
a coset representative, $\cS$:  $U \to  G$, defined to be 
a holomorphic isometry transformation $\cS(p)$:  $\cM \to \cM $ such that 
$$
\cS(p) \, p_0 =p~, \qquad \cS(p) \in G~,
$$ 
for any point $p \in U$. In other words, $\cS(p) $ maps the origin to $p$.
In local coordinates, $\cS(p) = \cS(\F, {\bar \F})$, and it acts on a generic point 
 $q \in U$ parametrized by complex variables $(\J^I , {\bar \J}^{\bar J} )$ as follows:
\bea
\J \to \J' = f (\J; \F, \bar \F )~, \qquad 
f (0; \F, \bar \F ) =\F~.
\eea
Now, we should point out that the holomorphic isometry transformations 
leave the equations  (\ref{asfem}) invariant.  
This means that applying the group transformation $\cS(\F, \bar \F )$ 
to $\U_0(\z)$, eq. (\ref{u-0}), gives
\bea 
\U_0(\z) ~\to ~ \U_*(\z) =  f (\U_0(\z) ; \F, \bar \F )
= f (\S_0 \,\z ; \F, \bar \F )~, \qquad 
 \U_*(0)=\F~. \label{sol-trans}
 \eea
Imposing the second initial condition in (\ref{geo3}),
\be
\S^I =\S^J_0 \, \frac{\pa 
}{\pa \J^J}
f^I (\J; \F, \bar \F ) \Big|_{\J=0} ~,
\label{generalizedlinear}
\ee
we are in a position to 
uniquely express $\S_0$ in terms of $\S$ and $\F$, $\bar \F$.
By construction, $\S$ is a complex linear superfield constrained   
as in (\ref{chiral+linear}). As to $\S_0$, it obeys a generalized 
linear constraint that follows from (\ref{generalizedlinear})
by requiring ${\bar D}^2 \S=0$.

Our consideration shows that $\U_*(\z)$ is independent of 
$\bar \S$,  i.e. $\U_*(\z) \equiv \U_*( \z; \F, {\bar \F}, \S )$,
for all Hermitian symmetric spaces.
The same conclusion also follows from the fact that,
in the case of Hermitian symmetric  spaces, 
the algebraic equations of motion are equivalent 
to the holomorphic geodesic equation
(with complex evolution parameter) \cite{GK1,GK2}
\bea
\frac{ {\rm d}^2 \U^I_* (\z) }{ {\rm d} \z^2 } + 
\G^I_{JK} \Big( \U_* (\z), \bar{\F} \Big)\,
\frac{ {\rm d} \U^J_* (\z) }{ {\rm d} \z } \,
\frac{ {\rm d} \U^K_* (\z) }{ {\rm d} \z }  =0 ~,
\eea
under the same initial conditions (\ref{geo3}).
Here $\G^I_{JK} 
( \F , \bar{\F} )$ are the Christoffel symbols for the  
K\"ahler metric $g_{I \bar J} ( \F , \bar{\F} ) = \pa_I 
\pa_ {\bar J}K ( \F , \bar{\F} )$.

A crucial element in the above scheme, for the Hermitian 
symmetric spaces, is the coset representative $\cS(\F, {\bar \F})$.
There is huge freedom in its choice, since 
it can always be replaced by  
$\cS(\F, {\bar \F}) \to \cS(\F, {\bar \F})\,h(\F, {\bar \F})$, 
with $h(\F, {\bar \F})$ an arbitrary function taking its values in the stability 
group $H$ of the origin, $\F=0$.
It is extremely important to use this freedom to choose 
the ``correct''  coset representative, since our final aim is to compute 
the tangent bundle action
\bea
 S_{{\rm tb}}[\F, \bar \F, \S, \bar \S] 
=  
\frac{1}{2\pi {\rm i}} \, \oint \frac{{\rm d}\z}{\z} \,  
 \int {\rm d}^8 z \, 
K \big( \U_* (\z), \breve{\U}_* (\z)  \big) ~.
\eea
With a complicated coset representative chosen, 
it will be practically impossible to do the contour integral on the right. 
In what follows, we will construct such a ``correct'' coset representatives 
for four series of compact  Hermitian 
symmetric spaces, and then extend the results to the non-compact 
case.

\sect{Algebraic setup: Compact case}

This section is devoted to the construction of coset representatives 
for the four series of irreducible compact  Hermitian 
symmetric spaces.

\subsection{The symmetric space ${\rm U}(n+m)/{\rm U}(n)\times {\rm U}(m)$}

The complex Grassmannian 
$G_{m,n+m}({\mathbb C})= {\rm U} (n+m) / {\rm U}(n) \times {\rm U}(m)$
is defined to be the space of $m$-planes 
through the origin in 
${\mathbb C}^{n+m}$.
Its elements can be considered to be  the equivalence classes
of  complex $(n+m)\times m$ matrices of rank $m$, 
\bea
x= (x^I{}_\b)
=\left(
\begin{array}{c}
x^i{}_\b \\
x_{\a\b} 
\end{array}
\right)
= \left(
\begin{array}{c}
\tilde{x} \\
\hat{x} 
\end{array}
\right)~, \qquad i=1,\ldots, n \qquad \a,\b=1,\ldots,m
\eea
defined modulo arbitrary transformations of the form
\bea
x~ \to ~ x\,g ~, \qquad g \in {\rm GL}(m, {\mathbb C})~. 
\label{gauge}
\eea
By applying such a transformation one can turn $x$ 
into a matrix $u$ 
constrained by
 \bea
&& \qquad 
u^\dagger u = \mathbbm{1} _m~.
\label{master}
\eea
In other words, the $m$ vectors $u_\b =(u^I{}_\b)$
form an orthonormal basis  on the $m$-plane.
Then, the `gauge' freedom (\ref{gauge}) reduces to 
\bea
u ~\to ~u\, g~, \qquad g \in {\rm U}(m)~.
\label{gauge2}
\eea

Consider the open domain in $G_{m,n+m}({\mathbb C})$
singled out  by the condition $\det \hat{u} \neq 0$
(an open coordinate chart in the Grassmann space).
Then, we can uniquely represent 
\bea
\hat{u} =s\,h~, \qquad 
 s^\dagger =s =(s_{\a \b})~,
\qquad h=(h^\a{}_\b ) \in {\rm U}(m)~,
\label{s-h}
\eea
with $s$ being a 
positive definite Hermitian matrix.
Equivalently we have
\bea
\hat{u} \, \hat{u} ^\dagger =s^2 ~, \qquad \hat{u} ^\dagger \hat{u}  
= h^{-1} s^2\,h~.
\label{decom}
\eea
Eq. (\ref{master}) becomes 
\bea
\tilde{u}^\dagger \tilde{u} + \hat{u}^\dagger \hat{u} 
= 
\tilde{u}^\dagger \tilde{u} +  h^{-1} s^2\,h= \mathbbm{1} _m~.
\label{master2}
\eea
It is worth pointing out that the `gauge' freedom (\ref{gauge2}) 
can be completely fixed by setting $h =\mathbbm{1} _m$.

Introduce an Hermitian $(n+m)\times (n+m)$ matrix $F(u)$,
\bea
F= \left(
\begin{array}{cc}
x\,\mathbbm{1}_n +  \tilde{u} \L \tilde{u}^\dagger ~  ~& ~\tilde{u} h^{-1}\\
h \,\tilde{u}{}^\dagger ~~& ~s 
\end{array}
\right)=F^\dagger~, \quad \L =\l(h^{-1}s\,h)=h^{-1}\l(s)\,h
 ~, \quad \L^\dagger =\L~.
\eea
Here $x$ is a real number, and $\l(s)$ some function.
We require $F(u)$ to be unitary,
$$
F^\dagger \,F =\mathbbm{1} _{n+m}~.
$$
This can be shown to hold if 
\be
\l (s) =- \frac{x\,\mathbbm{1} _m +s}{\mathbbm{1}_m -s^2}~, \qquad x^2=1~.
\ee
We have to choose
\be
x=-1\quad \longrightarrow \quad
\l (s) = \frac{\mathbbm{1} _m}{\mathbbm{1}_m +s}~, \label{lambda}
\ee
in order for $\l(s)$ to be well defined at $s_0=\mathbbm{1}_m$.
Here $s_0$ corresponds to
\bea
u_0
= \left(
\begin{array}{c}
\tilde{u}_0 \\
\hat{u}_0 
\end{array}
\right)=\left(
\begin{array}{c}
0 \\
\mathbbm{1} _m 
\end{array}
\right)~.
\label{u_0}
\eea
The crucial  property of $F(u)$ is that it 
maps $u_0$ to the equivalence class containing $u$:
\bea
F(u)\, u_0 =
 \left(
\begin{array}{c}
\tilde{u} h^{-1}\\
s 
\end{array}
\right) \sim
 \left(
\begin{array}{c}
\tilde{u} \\
\hat{u} 
\end{array}
\right) =u~.
\label{u_0-->u}
\eea
Let us point out that the  matrix $F(u)$ is  invariant under the `gauge' 
transformations (\ref{gauge2}).

It is useful to replace $F(u)$ by 
\bea
G(u) = F(u) 
\left(
\begin{array}{cc}
-\mathbbm{1} _n & 0 \\
0& \mathbbm{1} _m
\end{array}
\right)
= \left(
\begin{array}{cc}
\mathbbm{1}_n -  \tilde{u} h^{-1}\l(s)h \tilde{u}^\dagger ~  ~& ~\tilde{u} h^{-1}\\
-h \tilde{u}{}^\dagger ~~& ~s 
\end{array}
\right)~.
\label{cosetrep}
\eea
This matrix has the property that $G(u_0) =\mathbbm{1} _{n+m}$.
In the case $m=1$, the matrix $G$ is unimodular, 
$G(u) \in {\rm SU}(n+1)$, as can be checked using the identity
\begin{eqnarray}
\det \left(
\begin{array}{cc}
A & B \\
C& D
\end{array}
\right)= \det(A-B\,D^{-1}C) \,\det D~.\label{matrix}
\end{eqnarray}
Similar arguments can be used to show $G(u) \in {\rm SU}(n+m)$
in the case $n \geq m$.

Let us introduce local complex coordinates,
$\F =(\F^{i \a})$, in the Grassmann manifold 
\bea
u= \left(
\begin{array}{c}
\tilde{u} \\
\hat{u} 
\end{array}
\right)\quad \to \quad
\left(
\begin{array}{c}
\tilde{u}\, \hat{u}{}^{-1} \\
\mathbbm{1} _m
\end{array}
\right)
=\left(
\begin{array}{c}
\tilde{u}\, h^{-1} s^{-1} \\
\mathbbm{1} _m
\end{array}
\right) 
 \equiv
\left(
\begin{array}{c}
\F \\
\mathbbm{1} _m
\end{array}
\right) ~.
\label{complexcoor}
\eea
Eq. (\ref{master2}) is equivalent to
\bea
\F^\dagger \F + \mathbbm{1} _m = s^{-2}~. \label{s}
\eea
By construction, the variables $\F$ are invariant under 
the transformations (\ref{gauge2}).
Since the coset representative (\ref{cosetrep}) 
is also invariant under (\ref{gauge2}), 
the matrix elements of $G(u)$  
depend solely on  $\F$ 
and its conjugate:\footnote{In the case $m=1$, 
coset representative (\ref{cosetrep2})
reduces to that used in \cite{GK2} 
to derive the tangent bundle formulation 
for the $\cN=2$ supersymmetric sigma model (\ref{nact}) 
associated with $\mathbb{C}P^n$.}
\bea
G(u)=G(\F, \bar \F) = \left(
\begin{array}{cc}
\mathbbm{1}_n -  \F \,s\, \l (s) \,s\,\F^\dagger ~  ~& ~\F\,s\\
-s\, \F^\dagger ~~& ~s 
\end{array}
\right)~.
\label{cosetrep2}
\eea

Given an element of the isometry group, 
\bea
g= \left(
\begin{array}{cc}
A & B \\
C& D
\end{array}
\right) \,\in\, 
{\rm U}(n+m)~,
\eea
it
acts on a generic point (in the coordinate chart) of the Grassmann space
\bea
v= \left(
\begin{array}{c}
\tilde{v} \\
\hat{v} 
\end{array}
\right)
\sim
\left(
\begin{array}{c}
z \\
\mathbbm{1} _m
\end{array}
\right) 
\eea
by the holomorphic fractional linear transformation
\bea
\left(
\begin{array}{c}
z \\
\mathbbm{1} _m
\end{array}
\right) ~\to~\left(
\begin{array}{c}
z' \\
\mathbbm{1} _m
\end{array}
\right) ~,
\qquad 
z' = (A\, z +B)\, (C\,z+D)^{-1}.
\eea
Choosing here $g= G(\F,\bar \F )$ gives the action of 
the coset representative on the manifold.

Keeping in mind subsequent applications, let us describe 
a slightly different form for the coset representative
(\ref{cosetrep2}). Along with the matrix $s$, eq. (\ref{s}),
we can introduce
\bea
\underline{s}^2 =\frac{ \mathbbm{1} _n}
{\F \F^\dagger  + \mathbbm{1} _n }~,
 \label{u-s}
\eea
with the properties
\be
\F \,s =\underline{s} \,\F, 
\qquad 
\F^\dagger \,\underline{s}= s\,\F^\dagger ~.
\ee
Then, the coset representative (\ref{cosetrep2}) 
can be rewritten as follows:
\bea
G(\F, \bar \F) = \left(
\begin{array}{cc}
\underline{s} ~  ~& ~\F\,s\\
-\F^\dagger \, \underline{s}  ~~& ~s 
\end{array}
\right)~.
\label{cosetrep3}
\eea
This coset representative is well-known in the literature, 
see e.g. \cite{cosetrep}, and can be viewed as a generalization 
of the Wigner construction \cite{Wigner} used in his classification 
of the unitary representations of the Poincar\'e group.  

\subsection{The symmetric spaces ${\rm SO}(2n)/{\rm U}(n)$
and  ${\rm Sp}(n)/{\rm U}(n)$}

As is known \cite{Hua,CV} (see also \cite{HN} for a related discussion),
the Hermitian symmetric spaces ${\rm SO}(2n)/{\rm U}(n)$
and  ${\rm Sp}(n)/{\rm U}(n)$ can be realized as quadrics 
in the Grassmannian $G_{n,2n}({\mathbb C})$,
\be
x^{\rm T}{\bm J}_\e \,x =0~,
\label{quad}
\ee
where 
\bea
{\bm J}_\e = \left(
\begin{array}{cc}
0 & \mathbbm{1} _n \\
\epsilon\mathbbm{1} _n& 0
\end{array}
\right)~,
 ~~~~~\epsilon=
\left\{
\begin{array}{l}
 +~,~~~{\rm for}~~{\rm SO}(2n)/{\rm U}(n)~,\\
 -~,~~~{\rm for}~~ {\rm Sp}(n)/{\rm U}(n)~.
\end{array}
\right.
\label{j}
\eea
The submanifold (\ref{quad}) is invariant under the action 
of ${\rm O}(2n)$ for $\e=+1$, and  ${\rm Sp}(n)$ for $\e=-1$, 
with these groups realized as follows:
\bea
g= \left(
\begin{array}{cc}
A & B \\
C& D
\end{array}
\right) \in {\rm U}(2n)~, \qquad 
g^{\rm T} {\bm J}_\e \,g = {\bm J}_\e~.
\label{unort}
\eea
The two conditions (\ref{unort}) imply 
\bea
{\bar A} = D ~, \qquad {\bar C}=\epsilon B~,
\label{unort2}
\eea
with $\bar F$ denoting the complex conjugate of a matrix $F$.

In the complex local coordinates (\ref{complexcoor}), 
eq. (\ref{quad}) takes the form
\be
\F^{\rm T} +\e\, \F =0~. \label{const1}
\ee
In this case, eq. (\ref{s}) can be rewritten as
\bea
\mathbbm{1} _n + \F^\dagger \F =\mathbbm{1} _n - \epsilon\,{\bar \F}\,\F
= s^{-2}~,
\label{const0}
\eea
and for the matrix $\underline{s}$, eq. (\ref{u-s}), we obtain
\be
\underline{s} ={\bar s} ~.
\label{s-under-bar}
\ee 
${}$Finally,  the coset representative (\ref{cosetrep3}) 
turns into
\bea
&&G_\e (\F,\bar \F) = \left(
\begin{array}{cc}
\bar s & \F\,s ~~\\
\epsilon{\bar \F}\,{\bar s}~~& s
\end{array}
\right)~,
\quad  G_\e (\F,\bar \F) \in 
\left\{  
\begin{array}{l}
 {\rm U} (2n)  \cap {\rm SO} (2n)\,,
 \quad \e=1\,,
 \\
 {\rm U}(2n) \cap {\rm Sp}(n)\,,
 \quad \e=-1\,.
\end{array}
\right.
\label{coset-so}
\eea
Its crucial property is that it maps 
\bea
\left(
\begin{array}{c}
0 \\
\mathbbm{1} _n
\end{array}\right)~\to ~ 
G (\F ,\bar \F)
\left(
\begin{array}{c}
0 \\
\mathbbm{1} _n
\end{array} \right)
=\left(
\begin{array}{c}
\F \,s\\
s
\end{array}
\right)
\sim
\left(
\begin{array}{c}
\F \\
\mathbbm{1} _n
\end{array}
\right) ~.
\eea

\subsection{The symmetric space 
${\rm SO}(n+2)/{\rm SO}(n) \times {\rm SO}(2)$}
\label{1.3}

In accordance with \cite{CV,KN}, 
the Hermitian symmetric space\footnote{This space is irreducible 
for $n>2$.}
${\rm SO}(n+2)/{\rm SO}(n) \times {\rm SO}(2)$ 
is holomorphically equivalent to the complex quadric hypersurface 
$Q_n({\mathbb C}) $ in the projective space $ {\mathbb C}P^{n+1}$, 
and real-analytically isomorphic to the oriented 
Grassmann manifold $\tilde{G}_{2,n+2}({\mathbb R})$. 
Let us recall  the relevant  geometric constructions.

Consider the projective space 
$ {\mathbb C}P^{n+1} =G_{1,n+2}({\mathbb C})$.
Its elements are non-zero complex $(n+2)$-vectors,
\be
Z =(Z^I) \neq 0~, \qquad I=1, \dots , n+2~,  
\ee
defined modulo the equivalence relation 
\be
Z \sim \l \,Z~, \qquad \l \in {\mathbb C}^* = {\mathbb C}- \{0\}~.
\label{compquad-er1}
\ee
The complex quadric $Q_n({\mathbb C}) $ is the following 
hypersurface in $ {\mathbb C}P^{n+1}$:
\be
Q_n({\mathbb C})  =\Big\{ Z^{\rm T} Z 
= (Z^1)^2 + (Z^2)^2 +\dots + (Z^{n+2})^2 =0~, \qquad 
Z \in {\mathbb C}P^{n+1} \Big\}~.
\ee

Let $X$ and $Y$ be the real and imaginary parts of $Z$, respectively,
\be
Z= X+{\rm i}\, Y~.
\ee
On the quadric surface, $Z\in Q_n({\mathbb C}) $,
the real $(n+2)$-vectors $X$ and $Y$ obey the relations 
\bea
X^{\rm T} X= Y^{\rm T} Y \neq 0~, \qquad  
X^{\rm T} Y  =Y^{\rm T} X =0~.
\label{compor1}
\eea
In other words, considered as element of Euclidean space 
${\mathbb R}^{n+2}$, the non-zero vectors $X$ and $Y$ have the same length
and are orthogonal to each other. Therefore, they are linearly independent.
At this point, it is useful to introduce the $(n+2)\times 2$ matrix 
of rank 2:
\be
x = (x^I{}_\b )~, \qquad X = (x^I{}_1) ~, \quad 
Y= (x^I{}_2)~.
\ee 
Clearly, this matrix defines a two-plane through the origin 
in ${\mathbb R}^{n+2}$. Now, the relations (\ref{compor1}) 
can be rewritten as
\be 
x^{\rm T}x 
=\a\,
\mathbbm{1} _2~, \qquad \a=\hf \,{\rm tr} (x^{\rm T}x ) >0~, 
\label{compor2}
\ee
while the equivalence relation (\ref{compquad-er1}) turns into
\be
x \sim x\,g~, \qquad g \in {\mathbb R}^+ \times {\rm SO}(2)~,
\label{compquad-er2}
\ee
with ${\mathbb R}^+$ the multiplicative group of positive real numbers.

The real realization for  $Q_n({\mathbb C}) $ considered above, 
makes  obvious a relationship of this manifold to the 
oriented Grassmannian $\tilde{G}_{2,n+2}({\mathbb R})$ 
-- the space of oriented 
two-planes through the origin in ${\mathbb R}^{n+2}$.
The elements of $\tilde{G}_{2,n+2} (\mathbb R )$
can be identified with 
the equivalence classes of real $(n+2)\times 2 $ matrices of rank 2, 
\bea
x= (x^I{}_\b) =
\left(
\begin{array}{c}
x^i{}_\b \\
x_{\a \b} 
\end{array}
\right)
= \left(
\begin{array}{c}
\tilde{x} \\
\hat{x} 
\end{array}
\right)~, \qquad i=1,\ldots, n \qquad \a,\b=1,2
\label{x}
\eea
with respect to the equivalence relation
\bea
x~ \sim ~x\,g ~, \qquad g \in {\rm GL}^+(2, {\mathbb R})~. 
\label{er-quadric1-com}
\eea
Indeed, since $x$ is of rank 2, the matrix $x^{\rm T} x$ is positive 
definite. Then, by applying a transformation of the form
$x\to x\,g$, with $g \in {\rm GL}^+(2, {\mathbb R})$, 
one can always make $x$  obey eq. (\ref{compor2}).
Hence, we can identify  $Q_n({\mathbb C}) $ with 
 $\tilde{G}_{2,n+2} (\mathbb R )$.

Given  a matrix $x \in \tilde{G}_{2,n+2} (\mathbb R )$, 
its equivalence class contains a
matrix 
\bea
u= (u^I{}_\b) =
\left(
\begin{array}{c}
\tilde{u} \\
\hat{u} 
\end{array}
\right)~ 
\eea
constrained by
\bea
u^{\rm T} \,u
=\tilde{u}^{\rm T} \tilde{u} +\hat{u}^{\rm T} \hat{u} 
= \mathbbm{1} _2~.
\label{quadric-constr1-com}
\eea
Such a matrix defines an orthonormal basis on the two-plane 
chosen. In what follows,  
we deal with such matrices 
only, within each equivalence class, when studying various aspects 
of the Grassmann manifold.  
Under eq. (\ref{quadric-constr1-com}), the equivalence 
relation (\ref{er-quadric1-com}) reduces to 
\bea
u~ \sim~ u \,g ~, \qquad g \in {\rm SO}(2)~. 
\label{er-quadric2-com}
\eea

Let $U_{n+1,n+2}$ be 
the open domain in $\tilde{G}_{2,n+2}({\mathbb R})$
singled out  by the condition $\det \hat{u} \neq 0$
(an open coordinate chart in the Grassmann manifold).
It consists of the  two  components with empty intersection:
(i) $U_{n+1,n+2}^{(+)}$ in which   $\det \hat{u}> 0$; 
and (i) $U_{n+1,n+2}^{(-)}$ in which   $\det \hat{u}< 0$. 
They are mapped on each other, say,  by a rotation through angle $\p$ 
in the $(n,n+1)$ plane in ${\mathbb R}^{n+2}$. 
For our purposes, it will be sufficient to consider the 
chart $U_{n+1,n+2}^{(+)}$ only. 
Then,  we can  represent 
\bea
\hat{u} =s\,h~, \qquad s=s^{\rm T}= (s_{\a \b}) ~, 
\qquad h \in {\rm SO}(2)~,
\label{pd-o-com}
\eea
with $s$ positive definite.

Given a matrix $u \in U_{n+1,n+2}^{(+)}$, 
we can associate with it  the following 
${\rm SO}(n+2)$-transformation:
\begin{eqnarray}
 G(u)
= \left(
\begin{array}{cc}
\mathbbm{1}_n -  \tilde{u} h^{-1}
\l(s) h \tilde{u}^{\rm T} ~  ~& ~\tilde{u} h^{-1}\\
-h \,\tilde{u}{}^{\rm T} ~~& ~s 
\end{array}
\right)~, 
 \qquad
\l (s) = \frac{\mathbbm{1} _2}{\mathbbm{1}_2 +s}~.
\label{quadric-cosetrep-com}
\end{eqnarray}
The crucial  property of $G(u)$ is
\bea
G(u)\, u_0 =
 \left(
\begin{array}{c}
\tilde{u} h^{-1}\\
s 
\end{array}
\right)~, \qquad
u_0
= \left(
\begin{array}{c}
\tilde{u}_0 \\
\hat{u}_0 
\end{array}
\right)=\left(
\begin{array}{c}
0 \\
\mathbbm{1} _2
\end{array}
\right)~.
\eea

It is time to let the complex structure, which is intrinsically 
defined on $Q_n({\mathbb C}) $, enter the scene. 
Let us proceed by  introducing a complex $(n+2)$-vector 
of the form
\bea
w=(w^I) := (u^I{}_1 +{\rm i}\, u^I{}_2)
= \left(
\begin{array}{c}
\tilde{w} \\
\hat{w} 
\end{array}
\right)\equiv \left(
\begin{array}{c}
\vf \\
z 
\end{array}
\right)~, \qquad \vf = (\vf^i)~, 
\quad z = \left(
\begin{array}{c}
z_1 \\
z_2
\end{array}
\right)~.
\eea
Then, the  relations encoded in  (\ref{quadric-constr1-com}) 
are  rewritten as
\bea
w^{\rm T}w &=& \vf^{\rm T} \vf +(z_1)^2 +(z_2)^2 =0~, 
\label{quad1-com}
\\
w^\dagger w& =& \vf^\dagger \vf +|z_1|^2 +|z_2|^2 = 2~.
\label{quad2-com}
\eea
The equivalence relation  (\ref{er-quadric2-com}) turns into
\be
w ~\sim ~{\rm e}^{{\rm i}\, \s}w~, \qquad \s \in \mathbb R ~.
\label{er-quadric3-com}
\ee

In what follows, we choose the gauge condition 
$h=\mathbbm{1} _2$.
Recall that  $s =s^{\rm T}$ is positive definite, that is 
\bea
s = \left(
\begin{array}{cc}
s_{11}& s_{12} \\
s_{12} & s_{22}
\end{array}
\right)~, \qquad s_{11} >0~, \quad s_{22} >0~,
\quad s_{11}s_{22} - (s_{12})^2 >0~.
\eea
These results tell us that both the components of $z$,
\bea
 \left(
\begin{array}{c}
z_1 \\
z_2
\end{array}
\right)
:= \left(
\begin{array}{c}
(s\,h)_{11} +{\rm i} \,(s\,h)_{12} \\
(s\,h)_{12}  +{\rm i} \,(s\,h)_{22} 
\end{array}
\right)
= \left(
\begin{array}{c}
s_{11} +{\rm i} \,s_{12} \\
s_{12}  +{\rm i} \,s_{22} 
\end{array}
\right)~,
\eea
are non-vanishing, $z_{1,2} \neq 0$. If we further introduce new variables
\bea
z_\pm = z_1 \pm {\rm i} \,z_2~,
\eea
then one obtains\footnote{In deriving eq. (\ref{quad3-com}), 
we have fixed the gauge freedom  (\ref{er-quadric3-com})
by imposing the condition $h=\mathbbm{1} _2$. 
In general, the expression for $z_-$ is as follows:
$z_- = {\rm e}^{{\rm i}\,\s}(s_{11} +s_{22} )$, 
with $\s$ a real parameter.}
\bea
z_- =\overline{z_-}=s_{11} +s_{22} >0~, \qquad
\Big| \frac{z_+}{z_-}\Big|^2 <1~.
\label{quad3-com}
\eea
${}$Finally, introducing projective variables 
\bea
\F = \frac{\vf}{z_-}~, \qquad 
\r =\frac{z_+}{z_-}~,
\eea
the equations (\ref{quad1-com}) and (\ref{quad2-com}) turn into
\bea
\F^{\rm T}\F 
+\r &=&0~, 
\label{quad0} \\
2 \F^\dagger \F  +1 + |\r|^2 
&=&\frac{4}{(z_-)^2}~.
\label{z-com}
\eea
Eq. (\ref{quad3-com}) and (\ref{quad0}) tell us
\begin{eqnarray}
 |\Phi^{\rm T}\Phi|<1.
\end{eqnarray} 

Let us consider an isometry transformation $g \in {\rm SO}(n+2)$.
\bea
g= \left(
\begin{array}{cc}
A & B \\
C& D
\end{array}
\right) ~, \qquad 
g^{\rm T} \,g = \mathbbm{1}. 
\eea
The linear action $u \to u'=g u$ induces
the holomorphic fractional linear transformation
\bea
\F ~\to~\F' &=& \Big\{ (1,-{\rm i}) \Big( C\, \F + D \,\G(\F)
\Big) \Big\}^{-1}  
\Big\{ A \,\F + B\,\G(\F) \Big\} ~,
\non \\
\G (\F) &=& \hf \left(
\begin{array}{c}
1 -\F^{\rm T}\F \\
{\rm i} (1 + \F^{\rm T}\F) 
\end{array}
\right)~.
\label{IVholom-com}
\eea
Here we have used the fact that $z_-$ transforms as follows:
\be
\frac{z'_-}{z_-} = (1,-{\rm i}) \Big( C\, \F + D \,\G(\F)
\Big) \equiv {\rm e}^{\L(\F)}~.
\label{z--transfo-com}
\ee
Unlike $z_-$, its transform $z'_-$ is no longer real, 
but its phase is a gauge degree of freedom.

The K\"ahler potential \cite{Hua1,Hua2,MPS} is
\be
K(\F, \bar \F ) =\hf \ln \Big(1+2 \F^\dagger \F  +|\F^{\rm T}\F|^2 \Big)~.
\ee
Under the holomorphic isometry transformation (\ref{IVholom-com}), 
 it changes as 
\bea 
K(\F', \bar \F' ) =K(\F, \bar \F ) + \L(\F) + {\bar \L}(\bar \F )~, 
\eea
with $\L(\F) $ given in (\ref{z--transfo-com}).
This can be seen from the identity
\bea
1+2 \F'^\dagger \F'   
+|\r'|^2 
=\Big(1 + 2 \F^\dagger \F   
+|\r|^2 \Big)\,
\Big| \frac{z_-}{z'_-} \Big|^2~, \label{z-trans-quad}
\eea
in conjunction with eq. (\ref{z--transfo-com}).

Let us turn to the problem of expressing 
 the coset representative (\ref{quadric-cosetrep-com}) 
 in terms of the complex coordinates introduced above. 
In the gauge $h=\mathbbm{1} _2$
we can rewrite $G(u)$ as 
\bea
G(\F, \bar \F) = \left(
\begin{array}{cc}
\underline{s} ~  ~& ~\tilde{u} 
\\
-\tilde{u}{}^{\rm T} ~~& ~s 
\end{array}
\right)
\equiv
 \left(
\begin{array}{cc}
\cA ~  ~& ~\cB
\\
\cC ~~& ~ \cD
\end{array}
\right)
~, 
\label{quadric-cosetrep2-com}
\eea
where 
\be
\underline{s}^2 = \mathbbm{1} _n  - \tilde{u}\, \tilde{u}{}^{\rm T}~,
\qquad
s^2=\mathbbm{1}_2 -\tilde{u}^{\rm T}\tilde{u}~.
\ee
${}$For the matrix blocks in (\ref{quadric-cosetrep2-com}) we then get
\bea
\cA &=& \sqrt{
\mathbbm{1} _n - \frac{ z_-^2}{2} (\F\F^\dagger + {\bar \F} \F^{\rm T})
}~, \non \\
\cB&=& \hf z_- (\F\,, \,{\bar \F})\,\g~,
\qquad \cC = -\hf z_- \g^\dagger 
 \left(
\begin{array}{c}
\F^\dagger 
\\
\F^{\rm T}
\end{array}
\right)
~,
\non\\
\cD &=&{1 \over 2}\gamma^\dagger\sqrt{
\mathbbm{1} _2 - 
 \frac{ z_-^2}{2}\D
}\g\,,
\eea
where
\bea
\g=  \left(
\begin{array}{cr}
1~  & ~-{\rm i}
\\
1 ~& ~ {\rm i}
\end{array}
\right)~, \qquad 
\D = \left(
\begin{array}{cc}
\F^\dagger \F  ~& ~\overline{\F^{\rm T}\F}  \\
\F^{\rm T}\F~& ~ \F^\dagger \F
\end{array}
\right)~.
\eea

One can easily check that ${\cal D}$ can be rewritten as
\begin{eqnarray}
 {\cal D}={1 \over 4}z_-\gamma^\dagger F \gamma\,,\quad 
F=\left(
\begin{array}{cc}
 1 & -\overline{\Phi^{\rm T}\Phi}\\
 -\Phi^{\rm T}\Phi & 1
 \end{array}
\right)\,. \label{quad-eq1}
\end{eqnarray}
Furthermore it is useful to verify the following relations
\begin{eqnarray}
 \left(\begin{array}{c}
  \Phi^\dagger \\
  \Phi^{\rm T}
 \end{array}\right)
 {\cal A}
={1 \over 2}\gamma {\cal D}\gamma^\dagger 
 \left(\begin{array}{c}
  \Phi^\dagger \\
  \Phi^{\rm T}\end{array}\right)
={1 \over 2}z_-F
  \left(\begin{array}{c}
  \Phi^\dagger \\
  \Phi^{\rm T}\end{array}\right)\,, \label{quad-eq2} 
\end{eqnarray}
and 
\begin{eqnarray}
{\cal A}^{\rm T}{\cal A}=\mathbbm{1}_n-z_-^2(\Phi,\bar{\Phi})
  \left(\begin{array}{c}
  \Phi^\dagger \\
  \Phi^{\rm T}\end{array}\right)~. 
  \label{quad-eq3}
\end{eqnarray}

Equation (\ref{z-com}) gives the expression for $z_-$ 
in terms of $\F$ and its  conjugate. 
The isometry transformation $G(\F, \bar \F) \in {\rm SO}(n+2)$
 maps the origin, $\F_0=0$, to the point $\F$. 
On a generic point 
 $\U$ of the symmetric space, it acts by the rule:
\bea
\U ~\to~\U' &=& \Big\{ (1,-{\rm i}) \Big( \cC\, \U + \cD \,\G(\U)
\Big) \Big\}^{-1}  
\Big\{ \cA \,\U + \cB\,\G(\U) \Big\} ~,
\label{cr-cq}
\eea
with the two-vector $\G(\U)$ defined similarly to (\ref{IVholom-com}).

\sect{The (co)tangent bundle over 
${\rm U}(n+m)/{\rm U}(n)\times {\rm U}(m)$}

Here we apply the procedure described in section 2
to the case of Grassmann manifolds 
$G_{m,n+m}({\mathbb C})={\rm U}(n+m)/{\rm U}(n)\times {\rm U}(m)$.
In accordance with section 2, the tangent bundle action is 
\begin{eqnarray}
S=\frac{1 }{ 2\pi \rm i}\oint
\frac{ {\rm d} \zeta }{  \zeta}\int {\rm d}^8z \,
K(\Upsilon,\check{\Upsilon})~, 
\label{action}
\end{eqnarray}
with $K(\F, \bar \F)$ the K\"ahler potential.

In the case of the Grassmannian 
$G_{m,n+m}({\mathbb C})$,
the K\"ahler potential 
(see, e.g.  \cite{Chern}) is 
\be
K (\F , \F^\dagger )= \ln \det ( \mathbbm{1} _m + \F^\dagger \F) 
= \ln \det ( \mathbbm{1} _n + \F \F^\dagger )~,
\label{Kalpot}
\ee
where $\F= (\F^{i \a}) $ and $\F^\dagger = ({\bar \F}^{\bar \a \bar i})$,
with ${\bar \F}^{\bar \a \bar i} =\overline{\F^{i \a}}$. 
It will be assumed that the indices can be raised and lowered using 
the `flat metrics' $\d_{\a \bar \b} $ and $\d_{i \bar j}$, and their inverses, 
in particular ${\bar \F}^{\bar \a \bar i} = {\bar \F}_\a{}^{ \bar i}= {\bar \F}_{\a i}
={\bar \F}^{\bar \a}{}_i$.
The K\"ahler metric can readily be shown to be 
\bea
g_{i  \a ,\bar \b \bar j} &=&
\Big(\frac{\mathbbm{1} _m}{  \mathbbm{1} _m + \F^\dagger \F}\Big)_{\a \bar \b}
\left\{ \d_{\bar j i} - \Big(\F \, \frac{\mathbbm{1} _m}{  \mathbbm{1} _m + \F^\dagger \F} \,\F^\dagger
\Big)_{\bar j i} \right\} \non \\
&=& \Big(\frac{\mathbbm{1} _n}{  \mathbbm{1} _n + \F \F^\dagger }\Big)_{ \bar j i}
\left\{ \d_{\a \bar \b} - \Big(\F^\dagger \, \frac{\mathbbm{1} _n}{  \mathbbm{1} _n + \F \F^\dagger } \,\F
\Big)_{\a\bar \b} \right\} \non \\
&=& \Big(\frac{\mathbbm{1} _m}{  \mathbbm{1} _m + \F^\dagger \F}\Big)_{\a \bar \b}
 \Big(\frac{\mathbbm{1} _n}{  \mathbbm{1} _n + \F \F^\dagger }\Big)_{ \bar j i}~,
\label{metric}
\eea
where we have used the identities
\be
\frac{\mathbbm{1} _n}{  \mathbbm{1} _n + \F \F^\dagger }\,\F 
= \F \,\frac{\mathbbm{1} _m}{  \mathbbm{1} _m + \F^\dagger \F}~,
\qquad 
\F^\dagger \,\frac{\mathbbm{1} _n}{  \mathbbm{1} _n + \F \F^\dagger }
= \frac{\mathbbm{1} _m}{  \mathbbm{1} _m + \F^\dagger \F}\,\F^\dagger~. \label{identity1}
\ee

With the choice
\begin{eqnarray}
 K(\Upsilon,\breve{\Upsilon})
 =
 \ln \det ( \mathbbm{1} _m + \breve{\U}^{\rm T} \U) 
= \ln \det ( \mathbbm{1} _n + \U \breve{\U}^{\rm T} )
 \label{k-gr}
\end{eqnarray}
in action (\ref{action}), 
the equations of motion for the auxiliary superfields are 
\begin{eqnarray}
 \oint 
 \frac{{\rm d}\zeta }{ \zeta}\zeta^n 
 \left( \mathbbm{1}_m+\breve{\Upsilon}_*^{\rm T}\Upsilon_*
 \right)^{-1}\breve{\Upsilon}_*^{\rm T}=0\,,~~~n\ge 2~. 
 \label{eq-mot}
\end{eqnarray}
As explained in section \ref{2}, 
 one can easily check that (\ref{u-0}) solve the equations (\ref{eq-mot}).
According to (\ref{sol-trans}), we can obtain the solution at any point
 of the base manifold.
Acting by the coset representative (\ref{cosetrep}) on (\ref{u-0}), we
 obtain
\begin{eqnarray}
 \Upsilon_*=\left\{(\mathbbm{1} _n
-\Phi\lambda s^2 \Phi^\dagger)\S_0\z+\Phi s\right\}
 (-s \Phi^\dagger \S_0\z+s)^{-1}\,. \label{sol-gr}
\end{eqnarray}
${}$From here we read off the tangent vector at $\F$
\bea
\Sigma \equiv \frac{\partial \Upsilon_* }{ \partial \zeta}{\Bigg
  |}_{\zeta=0}
 =(\mathbbm{1} _n - \Phi\lambda s^2 \Phi^\dagger 
 + \Phi s \Phi^\dagger) \Sigma_0 s^{-1}
&=&(\mathbbm{1} _n + \Phi\lambda s \Phi^\dagger ) \Sigma_0 s^{-1} \non \\
&=& \underline{s}^{-1} \S_0 \, s^{-1}~.
\label{Sigma}
\eea
This result allows us to express $\S_0$ in terms of $\S$. 
Let us substitute the solution 
(\ref{sol-gr}) 
 into the potential (\ref{k-gr}). 
Then we have
\begin{eqnarray}
 K (\Upsilon_*,\breve{\Upsilon}_*)
 &=&\ln \det \Big(\mathbbm{1} _m +\breve{\Upsilon}_*^{\rm T} \Upsilon_* \Big) \nonumber \\
&=&\ln \det \Big(\mathbbm{1} _m +\Phi^\dagger \Phi-s^{-1}\Sigma_0^\dagger
\Sigma_0 s^{-1}\Big) 
\nonumber \\
&& -\ln \det \Big(\mathbbm{1} _m + {1 \over \zeta}s^{-1}\Sigma_0^\dagger \Phi s
\Big) - \ln\det  \Big(\mathbbm{1} _m-s\Phi^\dagger \Sigma_0
s^{-1}\zeta \Big)
\,. 
\label{k-gr2}
\end{eqnarray}
Here we have used eqs. (\ref{lambda}) and (\ref{s}), and their corollary
\be
\F s^2 \F^\dagger +(\mathbbm{1} _n - \Phi\lambda s^2 \Phi^\dagger )^2 =\mathbbm{1} _n\,. 
\ee
The expression in the last line of (\ref{k-gr2}) does not contribute to the action 
(\ref{action}) where the $\zeta$ integral only singles out the constant part, 
and it will not be written down explicitly in what follows.  
Now, eq. (\ref{Sigma})  implies $s^{-1}\S_0^\dagger = \S^\dagger \underline{s}$
and $\S_0 s^{-1} =  \underline{s} \S$, and hence
\bea
K (\U_* ,\breve{\Upsilon}_*) &=&
\ln \det \Big(\mathbbm{1} _m +\Phi^\dagger \Phi
-\Sigma^\dagger (\mathbbm{1} _n + \F \Phi^\dagger )^{-1}
\Sigma \Big) + \dots  
\non \\
&=& K(\F , \F^\dagger) 
+ \ln \det \Big( \mathbbm{1} _m
-(\mathbbm{1} _m +\Phi^\dagger \Phi)^{-1} \Sigma^\dagger (\mathbbm{1} _n + \F \Phi^\dagger )^{-1}
\Sigma \Big) + \dots  ,~~~~~     
\label{k-gr3}     
\eea
with $ K(\F , \F^\dagger) $  the K\"ahler potential of the base manifold, 
eq. (\ref{Kalpot}).  
Evaluated at $\U_*$ and $\breve{\Upsilon}_*$, the action (\ref{action})
turns into
the tangent bundle action
\bea
 S&=& \int {\rm d}^8z \left\{
 K(\F , \F^\dagger) 
+ \ln \det \Big( \mathbbm{1} _m
-(\mathbbm{1} _m +\Phi^\dagger \Phi)^{-1} \Sigma^\dagger (\mathbbm{1} _n + \F \Phi^\dagger )^{-1}
\Sigma \Big) \right\} 
\non \\
&=& \int {\rm d}^8z \left\{
 K(\F , \F^\dagger) 
+ \ln \det \Big( \mathbbm{1} _n
- (\mathbbm{1} _n + \F \Phi^\dagger )^{-1}
\Sigma \,
(\mathbbm{1} _m +\Phi^\dagger \Phi)^{-1} \Sigma^\dagger 
\Big) \right\} 
~. 
\label{action-gr}
\eea

It is not difficult to see that
\bea
I&=&\tr \ln  \Big( \mathbbm{1} _m
-(\mathbbm{1} _m +\Phi^\dagger \Phi)^{-1} \Sigma^\dagger (\mathbbm{1} _n + \F \Phi^\dagger )^{-1}
\Sigma \Big) \non \\
&=& \tr \ln  \Big( \mathbbm{1} _n
- (\mathbbm{1} _n + \F \Phi^\dagger )^{-1}
\Sigma \,
(\mathbbm{1} _m +\Phi^\dagger \Phi)^{-1} \Sigma^\dagger 
\Big)  
\eea
is actually a scalar field on an open domain of the zero section 
of the tangent bundle. By construction, $\S$ defines a holomorphic
tangent vector with world indices. Instead of using the coordinate 
basis, we can decompose tangent vectors with respect to the 
vielbein defined in (\ref{vielbein}), 
\bea
\S ~\to ~ \tilde{\S} = \underline{s} \,\S \,s~, \qquad 
\S^\dagger ~\to ~\tilde{\S}^\dagger = s\, \S^\dagger \,\underline{s}~.  
\label{tildeSigma}
\eea  
Then we readily obtain
\be 
I= \tr \ln  \Big( \mathbbm{1} _m
- \tilde{\S}^\dagger \tilde{\S}
\Big)
=  \tr \ln  \Big( \mathbbm{1} _n
- \tilde{\S} \tilde{\S}^\dagger 
\Big)~.
\ee
Our consideration shows that the tangent bundle action 
\bea
 S&=& \int {\rm d}^8z \left\{
 K(\F , \F^\dagger) 
+ \tr \ln  \Big( \mathbbm{1} _m
- \tilde{\S}^\dagger \tilde{\S}
\Big)
 \right\} 
\non \\ 
&=&
 \int {\rm d}^8z \left\{
 K(\F , \F^\dagger) 
+  \tr \ln  \Big( \mathbbm{1} _n
- \tilde{\S} \tilde{\S}^\dagger 
\Big) \right\} 
~. 
\label{action-gr-2}
\eea
is well-defined under the following covariant 
conditions\footnote{For an Hermitian
 matrix $H$, $H^\dagger =H$,
the notation $H> 0$ means that $H$ is positive definite.}
\bea
\tilde{\S}^\dagger \tilde{\S} < \mathbbm{1} _m
\quad \Longleftrightarrow \quad
 \tilde{\S} \tilde{\S}^\dagger < \mathbbm{1} _n~.
\label{Grassmann-global}
\eea
In appendix A, the curvature tensor of the Grassmannian is computed,
eq. (\ref{curvature-comp}). It follows from (\ref{curvature-comp})
that the Taylor expansion of (\ref{action-gr-2}) in powers 
of $\S$ and $\bar \S$ can be represented  in the universal form
(\ref{act-tab}).

By comparing the equations (\ref{Sigma}) and (\ref{tildeSigma}), 
one can see that $\tilde \S$ coincides with $\S_0$.
Nevertheless, here and below we prefer to use the ``tilde''
notation for (co)tangent vectors decomposed vectors with respect to the 
vielbein (\ref{vielbein}). 

Let us derive the cotangent bundle over the Grassmann manifold.
In order to obtain it, we need to dualize the complex
 linear superfields 
 $\S=(\Sigma^{i\alpha})$ 
  in (\ref{action-gr}) into chiral
 superfields 
  $\j=(\psi_{\alpha i})$
 forming the  components of a cotangent vector.
 To apply the relevant Legendre transformation,
 the action (\ref{action-gr}) is to be replaced by
 the following one
\begin{eqnarray}
 S&=& \int {\rm d}^8z \left\{
 K(\F , \F^\dagger) 
+ \tr \ln  \Big( \mathbbm{1} _m - 
 s  U^\dagger \,\underline{s}^2 U s \Big) +\hf \tr \,(U \j) 
 +\hf \tr \,(\j^\dagger U^\dagger  ) \right\}~,~~~
 \label{cot-gr0}
\end{eqnarray}
where $U=(U^{i\alpha})$ 
 is a complex unconstrained superfield.
By  construction, $U$ is a tangent vector at the point $\Phi$ of
 the base manifold.
Therefore $\psi$ is a one-form at the same point.
Varying the action with respect to $\j$ gives $U=\S$, 
and then one obtains the original action (\ref{action-gr}).
On the other hand, varying $U$ allows one to express 
$U$ in terms of $\F$, $\j$ and their conjugates, 
thus ending up with a dual formulation. 

In order to simplify further expressions and to make symmetry properties 
more transparent, 
it is useful to decompose the tangent and cotangent vectors 
with respect to the vielbein  (\ref{vielbein}), 
\bea
U ~\to ~ \tilde{U} = \underline{s} \,U \,s~, \qquad
\j ~\to ~ \tilde{\j} =s^{-1} \j\, \underline{s}^{-1} ~.
\eea  
Then, the action becomes
 \begin{eqnarray}
 S&=& \int {\rm d}^8z \left\{
 K(\F , \F^\dagger) 
+ \tr \ln  \Big( \mathbbm{1} _m - 
 \tilde{U}^\dagger \, \tilde{U}  \Big) + \hf \tr \,(\tilde{U} \tilde{\j}) 
 +\hf \tr \,(\tilde{\j}^\dagger \tilde{U}^\dagger  ) \right\}~.
 \label{cot-gr0-2}
\end{eqnarray}
We should point out that it is the variables $\j$ which are chiral, 
while their covariant counterparts, $\tilde \j$, 
obey a generalized chirality constraint.
To  eliminate the auxiliary fields $\tilde{U}$ and $\tilde{U}^\dagger$,
we consider their  equations of motion:
\bea
\hf \, \tilde{\j} = (\mathbbm{1}_m -\tilde{U}^\dagger \tilde{U} )^{-1}
\tilde{U}^\dagger = \tilde{U}^\dagger 
(\mathbbm{1}_n - \tilde{U} \tilde{U}^\dagger  )^{-1}~,
\label{j<---U}
\eea 
and the conjugate equation.
These  lead to 
$$
(\mathbbm{1}_n - \tilde{U} \tilde{U}^\dagger  )^{-1}
=\hf \Big(\mathbbm{1}_n  \pm 
\sqrt{\mathbbm{1}_n +\tilde{\j}^\dagger \tilde{\j}}\Big)~.
$$
We have to choose the ``plus'' solution in order to satisfy 
the requirement that $\tilde{\j}^\dagger \tilde{\j} \to0$ 
implies $\tilde{U} \tilde{U}^\dagger  \to 0$ and vice versa.
Thus
\bea
(\mathbbm{1}_n - \tilde{U} \tilde{U}^\dagger  )^{-1}
=\hf \Big(\mathbbm{1}_n  + 
\sqrt{\mathbbm{1}_n +\tilde{\j}^\dagger \tilde{\j}}\Big)~.
\eea
We also readily obtain 
\bea
\hf \,\tilde{U}\tilde{\j} =- \hf \Big(\mathbbm{1}_n  - 
\sqrt{\mathbbm{1}_n +\tilde{\j}^\dagger \tilde{\j}}\Big)~.
\eea

As a result the action (\ref{cot-gr0-2}) turns into
\begin{eqnarray}
 S&=& \int {\rm d}^8z \left\{
 K(\F , \F^\dagger) 
- \tr \ln  \Big( \mathbbm{1} _n 
+  \sqrt{\mathbbm{1}_n +\tilde{\j}^\dagger \tilde{\j}} \Big)
+\tr \,\sqrt{\mathbbm{1}_n +\tilde{\j}^\dagger \tilde{\j}} 
 \right\} \non \\
 &=& \int {\rm d}^8z \left\{
 K(\F , \F^\dagger) 
- \tr \ln  \Big( \mathbbm{1} _m 
+  \sqrt{\mathbbm{1}_m + \tilde{\j}\tilde{\j}^\dagger  } \Big)
+\tr \,\sqrt{\mathbbm{1}_m + \tilde{\j} \tilde{\j}^\dagger  } 
 \right\} ~.
 \label{cot-gr0-3}
\end{eqnarray}
This action defines the cotangent bundle formulation 
for the $\cN=2$ supersymmetric sigma model 
(\ref{action}) associated with 
the Grassmannian 
$G_{m,n+m}({\mathbb C})$.

\sect{The (co)tangent bundle over ${\rm SO}(2n)/{\rm U}(n)$ 
and  ${\rm Sp}(n)/{\rm U}(n)$}\label{so}

${}$For the Hermitian symmetric spaces
${\rm SO}(2n)/{\rm U}(n)$ and 
${\rm Sp}(n)/{\rm U}(n)$,
the K{\"a}hler potentials 
are known to be
\begin{eqnarray}
K(\Phi,\Phi^\dagger)=\ln\det(\mathbbm{1}_n+\Phi^\dagger\Phi)=\ln\det(\mathbbm{1}_n+\Phi\Phi^\dagger)~,
\label{kahler-so}
\end{eqnarray}
with the constraint (\ref{const1})
imposed on the variables 
$\Phi=(\Phi^{ij})$ and $\Phi^\dagger=(\bar{\Phi}^{\bar{j}\bar{i}})$,
where $i,j=1,\dots n$.
The K{\"a}hler metric can be read off as
\begin{eqnarray}
 g_{ik,\bar{l}\bar{j}}=\left({\mathbbm{1} _n \over {\bf
1}_n+\Phi^\dagger\Phi}\right)_{k\bar{l}}
\left({\mathbbm{1} _n \over \mathbbm{1} _n+\Phi\Phi^\dagger}\right)_{\bar{j}i}~,
\end{eqnarray}
where we have used (\ref{identity1}).

The ${\cal N}=2$ supersymmetric sigma model (\ref{action})
associated to (\ref{kahler-so}) is
generated by the Lagrangian (\ref{k-gr}), 
but now $\Upsilon$ has to obey the constraint
\begin{eqnarray}
\Upsilon^{\rm T}=-\epsilon \, \Upsilon~. 
\label{const-u-so}
\end{eqnarray}
The latter  follows from (\ref{const1}) and, in particular,  it requires
\begin{eqnarray}
 \Sigma^{\rm T}=-\epsilon \,\Sigma ~.
\label{tan-so}
\end{eqnarray}
Since the manifolds under consideration are imbedded 
into Grassmannians,
the equations of motion for the auxiliary superfields have 
the form (\ref{eq-mot}),  and their solution is given by 
(\ref{Sigma}). It follows from (\ref{Sigma}) 
\begin{eqnarray}
 \Sigma\equiv {\partial \Upsilon_* \over \partial \zeta}
 {\Big |}_{\zeta=0}
%=\bar{s}\Sigma_0 s^{-1}-\epsilon \Phi 
 = \bar{s}^{\,-1}\Sigma_0\, s^{-1}  ~,
\label{sigma-so}
\end{eqnarray}
where we have used the identity (\ref{s-under-bar})
which holds for the manifolds ${\rm SO}(2n)/{\rm U}(n)$ 
and  ${\rm Sp}(n)/{\rm U}(n)$.
Requiring $\Sigma^{\rm T}_0=-\epsilon \, \Sigma_0$,
it then follows from (\ref{sigma-so}) 
that $\S$ indeed obeys the algebraic constraint (\ref{tan-so}).

Now, it is obvious that 
the tangent bundle action for the symmetric spaces 
SO$(2n)$/U$(n)$ and Sp$(n)$/U$(n)$
is given by eq. (\ref{action-gr}) with $m=n$.

To derive the cotangent bundle formulation, 
we can again use the first-order action (\ref{cot-gr0})
with $m=n$ 
in which, however, the tangent $U=(U^{ij})$
and cotangent $\j=(\j_{ij})$ variables must 
obey the algebraic conditions
\bea
U^{\rm T}=-\epsilon \,U ~, \qquad 
\j^{\rm T}=-\epsilon \, \j ~.
\label{Uj-symmetry}
\eea
The equations of motion for $U$ and $U^\dagger$ should 
respect these symmetry conditions.
At first sight, one could then think that  the equation  (\ref{j<---U}) 
should be modified in order to accommodate these conditions.
${}$Fortunately, the right-hand side of (\ref{j<---U}) automatically enjoys
the desirable symmetry conditions, 
\bea
\tilde{\j} \equiv 2\,(\mathbbm{1}_n -\tilde{U}^\dagger \tilde{U} )^{-1}
\tilde{U}^\dagger = 2\,\tilde{U}^\dagger 
(\mathbbm{1}_n - \tilde{U} \tilde{U}^\dagger  )^{-1} 
=-\epsilon \, \tilde{\j}{}^{\rm T} 
\label{j<---U2}
\eea 
provided $U$ is chosen to obey 
the corresponding algebraic constraint in  (\ref{Uj-symmetry}), 
$U^{\rm T}=-\epsilon \,U$, and therefore 
$\tilde{U}{}^{\rm T}=-\epsilon \,\tilde{U}$.
As a result, all the steps implemented below eq.  (\ref{j<---U}) 
to derive (\ref{cot-gr0-3}),
remains valid in the case under consideration.

We conclude that  the action (\ref{cot-gr0-3}) with $m=n$
 defines the cotangent bundle formulation 
for the $\cN=2$ supersymmetric sigma model 
(\ref{action}) associated with 
the symmetric spaces
SO$(2n)$/U$(n)$ and Sp$(n)$/U$(n)$.

\sect{The (co)tangent bundle over 
${\rm SO}(n+2)/{\rm SO}(n) \times {\rm SO}(2)$}
${}$For the $\cN=2$ supersymmetric sigma model (\ref{nact})
associated with the Hermitian symmetric  space ${\rm SO}(n+2)/{\rm SO}(n) \times {\rm SO}(2)$,
the (co)tangent bundle formulations have  
 been studied in Ref. \cite{AN}.
The approach of \cite{AN} was based on implementing the following steps:
(i) construct a coset representative in the case $n=2$;
(ii) apply it to construct the corresponding  solution $\U_*(\z)$;
(iii) make use of the latter in order to guess the explicit form of $\U_* (\z)$ for 
$n>2$.

In this section, we are going to address the same problem by different means --
the coset representative  (\ref{quadric-cosetrep2-com}) allows us to carry out 
the scheme described in section 2 for general $n$. As is explained below,
this leads to somewhat different conclusions  for the cotangent bundle formulation.

The K{\"a}hler potential of the quadric surface 
\cite{Hua1,Hua2,MPS}
 is\footnote{Taking the normalization $r\rightarrow r/\sqrt{2}$ 
 in Ref. \cite{AN} gives our normalization.}
 \begin{eqnarray}
 K(\Phi,\bar{\Phi})={1 \over 2}\ln{4 \over z^2_-}, \quad
 z_-^2={4 \over 1 + 2\Phi^\dagger \Phi+|\Phi^{\rm T}\Phi|^2}\,.
\end{eqnarray}
Its ${\cal N}=2$ extension is given by
\begin{eqnarray}
 K(\Upsilon,\breve{\Upsilon})={1 \over 2}
\ln\left(1 + 2\breve{\Upsilon}^{\rm T} \Upsilon + {\Upsilon}^{\rm
     T}\Upsilon \breve{\Upsilon}^{\rm T}\breve{\Upsilon} \right)\,. \label{kahler-quad}
\end{eqnarray}
${}$From here we read off
the equations of motion for the auxiliary superfields 
\begin{eqnarray}
0=
\oint {{\rm d}\zeta \over \zeta}\zeta^n
 \frac{ \breve{\Upsilon}
 + \Upsilon
 \,\breve{\U}^{\rm T} \breve{\U}}
 {  1 + 2 \breve{\Upsilon}^{\rm T} \Upsilon
+\Upsilon^{\rm T} \Upsilon
\,\breve{\U}^{\rm T} \breve{\U} }~,
~~~~n\ge 2\,. \label{auxiliary-q}
\end{eqnarray}
The solution to (\ref{auxiliary-q}) is obtained 
from (\ref{cr-cq}) by replacing
 $\Upsilon\rightarrow \Sigma_0\z$  
 and $\Upsilon^\prime \rightarrow \Upsilon_*(\z)$,
 with $\S_0$ a tangent vector at $\F=0$.  
Then, we have
\begin{eqnarray}
 \U_*(\z) = \frac{ \F +(2/ z_-)\,   
{\cal A} \Sigma_0 \,\zeta -
\bar{\Phi} \,\Sigma_0^{\rm T}\Sigma_0\, \zeta^2
}
{1 - 2 \F^\dagger \S_0\, \z + 
\overline{\F^{\rm T}\F}\, \S_0^{\rm T} \S_0 \, \z^2 }~. 
\label{sol0-quad}
\end{eqnarray}
${}$For the tangent vector $\Sigma$ at $\F$, we then obtain 
\begin{eqnarray}
 \Sigma\equiv {\partial \Upsilon_* \over \partial \zeta}{\Bigg |}_{\zeta=0}
 =\frac{2}{z_-}{\cal A}\Sigma_0+2\Phi(\Phi^\dagger\Sigma_0)\,. \label{sigma-quad}
\end{eqnarray}

Using eq.
(\ref{sigma-quad}), 
 one can express $\Sigma_0$ via $\Sigma$ with the aid of
relations (\ref{quad-eq1}) and (\ref{quad-eq2}).
One derives 
\begin{eqnarray}
\F^\dagger \S_0 &=& \frac{1}{4}z_-^2 \Big\{ 
\F^\dagger \S + \overline{\F^{\rm T}\F}\,\F^{\rm T}\S \Big\}~, \non \\
\F^{\rm T} \S_0 &=& \frac{1}{4}z_-^2 \Big\{ (1+2|\F|^2 )\,\F^{\rm T} \S
- \F^{\rm T}\F\,\F^\dagger \S \Big\}~, \non \\
\S_0^{\rm T} \S_0 &=& \frac{1}{4}z_-^2 \S^{\rm T} \S~, 
\label{S-0-->S-quad} \\
\S_0^{\dagger} \S_0 &=& \frac{1}{4}z_-^2 \S^{\dagger} \S  
+ \frac{1}{8}z_-^4
\Big\{ (1+2|\F|^2 )\,\F^{\rm T} \S \, \S^\dagger {\bar \F} \non \\
&& \qquad \qquad 
-\overline{\F^{\rm T}\F}\,\F^{\rm T}\S \, \S^\dagger \F 
- \F^{\rm T}\F\,\F^\dagger \S \, \S^\dagger {\bar \F} 
- \F^\dagger \S \, \S^\dagger \F \Big\} ~. \non 
\end{eqnarray} 
These relations lead to the final form for the solution $\Upsilon_*$:
\bea
\U_*(\z) = \frac{ 
\F +  \zeta \S - (z_-^2/2) \Big\{ \z \F(  \F^\dagger \S 
+ \overline{\F^{\rm T}\F}\,\F^{\rm T}\S ) 
+\hf \z^2 {\bar \F}\, \S^{\rm T} \S \Big\}
}
{1 - \hf z_-^2  \Big\{ \z ( \F^\dagger \S 
+ \overline{\F^{\rm T}\F}\,\F^{\rm T}\S ) 
- \hf \z^2 \overline{\Phi^{\rm T}\Phi} \S^{\rm T} \S \Big\} }~.
 \label{sol-quad}
\eea

Now let us turn to calculating the Lagrangian
\begin{eqnarray}
 L={1 \over 2\pi \rm  i}\oint {{\rm d}\zeta \over
  \zeta}K(\Upsilon_*,\breve{\Upsilon}_*)\,. \label{lag}
\end{eqnarray}
Considerations similar to those used to derive 
eq. (\ref{z-trans-quad}) 
give
\begin{eqnarray}
 K(\U_*,\breve{\U}_*)&=& K(\F , \bar \F) 
+\hf \ln \Big( 1 - 2 \S_0^\dagger \S_0 + |\S^{\rm T}_0\S_0|^2 \Big)
-\hf \ln x(\z) - \hf \ln \breve{x} (\z)~, ~~~\\
x(\z) &=& 1 - 2 \F^\dagger \S_0\, \z + 
\overline{\F^{\rm T}\F}\, \S_0^{\rm T} \S_0 \, \z^2 ~.
\non
\end{eqnarray}
The last two terms 
in the expression for $ K(\U_*,\breve{\U}_*)$
do not contribute to the Lagrangian upon the
 integration over $\zeta$. 
Using the third and forth formulas in (\ref{S-0-->S-quad}), we obtain
\begin{eqnarray}
L & = &  K(\F , \bar \F) 
 +\hf \ln \Big( 1 - \frac{1}{2}z_-^2 \S^{\dagger} \S  
 - \frac{1}{4}z_-^4
 \Big\{ (1+2|\F|^2 )\,\F^{\rm T} \S \, \S^\dagger {\bar \F} \non \\
&& 
-\overline{\F^{\rm T}\F}\,\F^{\rm T}\S \, \S^\dagger \F 
- \F^{\rm T}\F\,\F^\dagger \S \, \S^\dagger {\bar \F} 
- \F^\dagger \S \, \S^\dagger \F
+\frac{1}{16} z_-^4 |\S^{\rm T} \S|^2
 \Big\} \Big)~.
\end{eqnarray}

The Lagrangian  can be rewritten in a geometric form.
In order to do that, we use the metric of the compact quadric surface 
\begin{eqnarray}
 g_{i\bar{j}}=
   \frac{1}{4}z_-^2
 \delta_{i\bar{j}}
 +\frac{1}{8}z_-^4
 \left\{
 \Phi^i\bar{\Phi}^{\bar{j}} (1+2|\F|^2) 
- \bar{\Phi}^{\bar{i}} \F^j 
- \Phi^i\Phi^{j}(\overline{\Phi^{\rm T}\Phi})
- \bar{\Phi}^{\bar{i}}\bar{\Phi}^{\bar{j}}(\Phi^{\rm T}\Phi) 
\right\}\,.
\end{eqnarray}
From this we have
\begin{eqnarray}
g_{i\bar{j}}\Sigma^i\bar{\Sigma}^{\bar{j}}=
  \frac{1}{4}z_-^2
 |\Sigma|^2
&+& 
 \frac{1}{8}z_-^4
\Big\{
  (1 + 2|\F|^2) 
 |\Phi^{\rm T}\Sigma|^2
 -|\Phi^\dagger \Sigma|^2 
\nonumber \\
 &-& (\Phi^{\rm  T}\Phi)
 (\Phi^\dagger\bar{\Sigma})(\Phi^\dagger\Sigma)
  - (\overline{\Phi^{\rm T}\Phi}) 
 (\Phi^{\rm T}\Sigma)(\Phi^{\rm T}\bar{\Sigma})
  \Big\}~.
\end{eqnarray}
Thus 
\bea
L & = &  K(\F , \bar \F) 
+\hf \ln \Big( 1 - 2 g_{i\bar{j}}\Sigma^i\bar{\Sigma}^{\bar{j}}
+ \frac{1}{16}z_-^4
|\S^{\rm T} \S|^2 \Big)~.
\label{LL-quad}
\eea
Here the second term is a scalar field on the tangent bundle.
Therefore,
the combination $z_-^4|\S^{\rm T} \S|^2 $ must be a scalar
constructed in terms of  the tangent vector $\S^i$ and its conjugate, the metric
 $g_{i\bar{j}}$ and the Riemann curvature
 $R_{i\bar{j}k\bar{l}}\equiv
 \partial_k\bar{\partial}_{\bar{l}}g_{i\bar{j}}
 -g^{m\bar{n}}\partial_{m}g_{i\bar{j}}\bar{\partial}_{\bar{n}}g_{k\bar{l}}$, 
 with $\partial_i=\partial / \partial \Phi^i$.
It is sufficient to determine such an expression
 at any given point of the base manifold, say at $\F=0$, 
 since the base manifold is a symmetric space.
This gives 
\begin{eqnarray}
2(g_{i\bar{j}}\Sigma^i\bar{\Sigma}^{\bar{j}})^2 + \hf R_{i\bar{j}k\bar{l}}
 \Sigma^i\bar{\Sigma}^{\bar{j}}\Sigma^k\bar{\Sigma}^{\bar{l}}
&=&
 \frac{1}{16}z_-^4
|\S^{\rm T} \S|^2 ~.
\end{eqnarray}
As a result, we arrive at 
the tangent bundle action \cite{AN}
\begin{eqnarray}
 S=\int {\rm d}^8z\left\{ K(\Phi,\bar{\Phi})
+\hf\ln \Big( 1 - 2g_{i\bar{j}}\Sigma^i\bar{\Sigma}^{\bar{j}}
+2 (g_{i\bar{j}}\Sigma^i\bar{\Sigma}^{\bar{j}})^2
+\hf R_{i\bar{j}k\bar{l}}
 \Sigma^i\bar{\Sigma}^{\bar{j}}\Sigma^k\bar{\Sigma}^{\bar{l}} \Big)
\right\}~.~~~~ 
\label{action-quad}
\end{eqnarray}

Note that the tangent vector in (\ref{action-quad})
should be  constrained as follows:
\begin{eqnarray}
 1 - 2g_{i\bar{j}}\Sigma^i\bar{\Sigma}^{\bar{j}}
+2 (g_{i\bar{j}}\Sigma^i\bar{\Sigma}^{\bar{j}})^2
+\hf R_{i\bar{j}k\bar{l}}
 \Sigma^i\bar{\Sigma}^{\bar{j}}\Sigma^k\bar{\Sigma}^{\bar{l}}&>&0~,
\non \\
g_{i\bar{j}}\Sigma^i\bar{\Sigma}^{\bar{j}} &<& 1~.
\label{tb-domain-quadric}
\end{eqnarray} 
To make the action (\ref{action-quad}) well-defined, 
we actually need only the first constraint in (\ref{tb-domain-quadric}).
The latter can be shown to imply 
$g_{i\bar{j}}\Sigma^i\bar{\Sigma}^{\bar{j}} \neq 1$, 
and therefore its space of solutions 
consists of  two connected components.
The second constraint in (\ref{tb-domain-quadric})
picks up one of the components.

${}$Finally, it remains to 
dualize 
the tangent bundle action (\ref{action-quad}) in order to 
generate  the cotangent bundle formulation.
The derivation is very similar to that performed for the non-compact quadric
 surface in section \ref{4} and Appendix \ref{noncq-hcp-der}.
Here we only give the result: 
\begin{eqnarray}
S=\int {\rm d}^8z {\Bigg \{}K(\Phi,\bar{\Phi})
 - \hf \ln \Big(  \L
&+& \sqrt{2(\L + g^{\bar{i}\,j}\bar{\psi}_{\bar{i}} \j_j)\,
}
\Big)
+\frac{1}{4}\left(\L+\sqrt{2(\L+g^{\bar{i}j}\bar{\psi}_{\bar{i}}\psi_j)}\right)
\non \\
&+& \frac{1}{2}\frac{
  \big(g^{\bar{i}\,j}\bar{\psi}_{\bar{i}} \j_j \big)^2
+\frac{1}{4} R^{\bar{i}j\bar{k}l}\bar{\psi}_{\bar{i}}\psi_{j}\bar{\psi}_{\bar{k}}\psi_l 
}
{\L
+ \sqrt{ 2\big(\L + g^{\bar{i}\,j}\bar{\psi}_{\bar{i}} \j_j \big)}\,
}{\Bigg \}}~,
\label{cq-hcp}
\eea
where 
\bea 
\L = 1 +\sqrt{ 
1 + 2 g^{\bar{i}\,j}\bar{\psi}_{\bar{i}} \j_j 
+2  (g^{\bar{i}\,j}\bar{\psi}_{\bar{i}} \j_j )^2
+\hf R^{\bar{i}j\bar{k}l}\bar{\psi}_{\bar{i}}\psi_{j}\bar{\psi}_{\bar{k}}\psi_l 
}~.
\eea

The dualization was also
 studied in Ref. \cite{AN}.
The derivation was performed in the case $n=2$ case and it was claimed that the
 cotangent bundle action should  be valid for general $n$ case
 since it is written in  geometric terms.
In the present paper,  the cotangent bundle action is derived for 
general $n$, and its explicit form 
 is different from that obtained in Ref. \cite{AN}.

\sect{Algebraic setup: Non-compact case}

This section is devoted to the construction of coset representatives 
for the four series of non-compact  Hermitian 
symmetric spaces.

\subsection{The symmetric space ${\rm U}(n,m)/{\rm U}(n)\times {\rm U}(m)$}

The non-compact Hermitian symmetric space 
${\rm U} (n,m) / {\rm U}(n) \times {\rm U}(m)$
can be identified with an  open subset of  
$G_{m,n+m}({\mathbb C}) $.
This subset consists  of those $m$-planes in ${\mathbb C}^{n+m}$
which obey the equation
\bea
x^\dagger\, \O \,x >0 ~, 
\qquad 
\O =   \left(
\begin{array}{cc}
- \mathbbm{1} _n & 0\\
0 & \mathbbm{1} _m
\end{array}
\right)~.
\label{bounddomain}
\eea
Here $x$  is a complex $(n+m)\times m$ matrix of rank $m$, 
\bea
x= (x^I{}_\b)
=\left(
\begin{array}{c}
x^i{}_\b \\
x_{\a\b} 
\end{array}
\right)
= \left(
\begin{array}{c}
\tilde{x} \\
\hat{x} 
\end{array}
\right)~, \qquad i=1,\ldots, n \qquad \a,\b=1,\ldots,m
\eea
which is defined modulo arbitrary transformations of the form
\bea
x~ \to ~ x\,g ~, \qquad g \in {\rm GL}(m, {\mathbb C})~. 
\label{nc-gauge}
\eea
One of the consequences of
eq. (\ref{bounddomain}) is that  $\hat{x} \in {\rm GL}(m, \mathbb C )$. 

By applying a  transformation of the form (\ref{nc-gauge}), 
one can turn $x$ into a matrix $u$
under the equation
\bea
u^\dagger\, \O \,u = 
\hat{u}^\dagger \hat{u}- 
\tilde{u}^\dagger \tilde{u} 
= \mathbbm{1} _m
\quad \longrightarrow \quad 
\det \hat{u} \neq 0~.
\label{nc-master}
\eea
With such a choice, the `gauge' freedom  (\ref{nc-gauge}) reduces to 
\bea
u~ \to~ u\,g ~, \qquad g \in {\rm U}(m)~. 
\label{nc-gauge2}
\eea
We further  represent $\hat{u}$ 
according to  eq. (\ref{s-h}), 
with $s$ being a uniquely chosen positive definite Hermitian matrix.
Then, eq. (\ref{nc-master}) becomes
\bea
\hat{u}^\dagger \hat{u} - \tilde{u}^\dagger \tilde{u} 
= h^{-1} s^2\,h
-\tilde{u}^\dagger \tilde{u} 
= \mathbbm{1} _m~.
\label{nc-master2}
\eea

Let us introduce 
\bea
G(u) = \left(
\begin{array}{cc}
\mathbbm{1}_n +  \tilde{u} h^{-1}
\l(s) h \tilde{u}^\dagger ~  ~& ~\tilde{u} h^{-1}\\
h \,\tilde{u}{}^\dagger ~~& ~s 
\end{array}
\right)~, \qquad
\l (s) = \frac{\mathbbm{1} _m}{\mathbbm{1}_m +s}~.
\label{nc-cosetrep}
\eea
The matrix $G(u)$ has the properties
\bea
G^\dagger (u) =G (u)~, 
\qquad G(u) \in {\rm SU}(n,m)~.
\eea
Another crucial feature of $G(u)$ is that it 
enjoys the property (\ref{u_0-->u}), with 
$u_0$ defined in (\ref{u_0}).
In other words,  $G(u)$ is a global coset representative 
for ${\rm U} (n,m) / {\rm U}(n) \times {\rm U}(m)$.

We can introduce global complex coordinates on 
${\rm U} (n,m) / {\rm U}(n) \times {\rm U}(m)$
\bea
u= \left(
\begin{array}{c}
\tilde{u} \\
\hat{u} 
\end{array}
\right)\quad \to \quad
\left(
\begin{array}{c}
\tilde{u}\, \hat{u}{}^{-1} \\
\mathbbm{1} _m
\end{array}
\right)
=\left(
\begin{array}{c}
\tilde{u}\, h^{-1} s^{-1} \\
\mathbbm{1} _m
\end{array}
\right) 
 \equiv
\left(
\begin{array}{c}
\F \\
\mathbbm{1} _m
\end{array}
\right) ~.
\eea
Eq. (\ref{nc-master2}) is equivalent to
\bea
 \mathbbm{1} _m -\F^\dagger \F 
= s^{-2}~. \label{nc-s}
\eea
Since $s>0$, we equivalently have 
\bea
\F^\dagger \F < \mathbbm{1} _m ~.
\eea
This relation defines a classical bounded symmetric domain
\cite{Cartan,Hua}.
Now, the coset representative (\ref{nc-cosetrep})
turns into
\bea
G(\F,\bar \F ) = \left(
\begin{array}{cc}
\mathbbm{1}_n +  \F \,s\, \l (s) \,s\,\F^\dagger ~  ~& ~\F\,s\\
s\, \F^\dagger ~~& ~s 
\end{array}
\right)~.
\label{nc-cosetrep2}
\eea

The coset representative obtained
can also be rewritten in the form
\bea
G(\F,\bar \F ) = \left(
\begin{array}{cc}
\underline{s}
~  ~& ~\F\,s\\
\F^\dagger \,\underline{s}~~& ~s 
\end{array}
\right)~,
\label{nc-cosetrep3}
\eea
where the matrices $s$ and $\underline{s}$ are defined as
\bea
s^2 =
\frac{ \mathbbm{1} _n}
{ \mathbbm{1} _n- \F^\dagger  \F}~, \qquad
\underline{s}^2 =
\frac{ \mathbbm{1} _n}
{ \mathbbm{1} _n -\F \F^\dagger  }~, \qquad 
\F s^2 = \underline{s}^2 \F~, \qquad
s^2 \F^\dagger= \F^\dagger \underline{s}^2 ~.
\label{nc-s2}
\eea

\subsection{The symmetric spaces ${\rm SO}^*(2n)/{\rm U}(n)$ and 
${\rm Sp}(n,{\mathbb R})/{\rm U}(n)$  }
${}$Following \cite{Hua}, the Hermitian symmetric spaces
${\rm SO}^*(2n)/{\rm U}(n)$ and  ${\rm Sp}(n,{\mathbb R})/{\rm U}(n)$  
can be identified with  special open domains
in $G_{n,2n}({\mathbb C}) $
consisting  of those  $n$-planes 
in ${\mathbb C}^{2n}$
which obey the constraints
\bea
x^\dagger\, \O \,x >0 ~, \qquad 
x^{\rm T}  \,{\bm J}_\e \,x =0~,
\qquad 
\O =   \left(
\begin{array}{cc}
- \mathbbm{1} _n & 0\\
0 & \mathbbm{1}_n
\end{array}
\right)~.
\label{bounddomain2}
\eea
Here the matrix $\bm J_\e$ is defined in (\ref{j}), with $\e=+1$ 
corresponding to ${\rm SO}^*(2n)$, 
and $\e=-1$ to  ${\rm Sp}(n,{\mathbb R})$.
It is pertinent to  recall the definition of the groups 
${\rm SO}^*(2n)$ and  ${\rm Sp}(n,{\mathbb R})$
(see, e.g. \cite{Knapp}):
\bea
{\rm G}= \Big\{ g \in {\rm SU}(n,n)~, 
\quad 
g^{\rm T} {\bm J}_\e \,g = {\bm J}_\e \Big\} ~, \qquad 
\e= \left\{ 
\begin{array}{cc}
+1~,  ~~~& {\rm for} ~{\rm SO}^*(2n)~,\\
-1~ , ~ ~~& {\rm for}~ {\rm Sp}(n,{\mathbb R})~.
\end{array}
\right. 
\eea

We can introduce global complex coordinates on the manifold,
\bea 
X ~ \sim ~
\left(
\begin{array}{c}
\F \\
\mathbbm{1} _n
\end{array}
\right) ~, \qquad \F^{\rm T} +\e \F =0~,
\label{nc-sa}
\eea
with the $ n\times n$ matrix $\F$ constrained as
\bea
\F^\dagger \F < \mathbbm{1} _n ~.
\eea

In the case $n=m$, 
if we choose 
$\F $  in (\ref{nc-cosetrep3}) to be 
antisymmetric or symmetric, 
$  \F^{\rm T} =-\e \F $, then $\underline{s} = \bar s$, 
with $\bar s$ the complex conjugate of $s$.
As a result, the coset representative (\ref{nc-cosetrep3})
becomes
\bea
G_\e(\F,\bar \F ) = \left(
\begin{array}{cc}
{\bar s}
~  ~& ~\F\,s\\
-\e{\bar \F} \,{\bar s}~~& ~s 
\end{array}
\right)~, \qquad 
G_\e(\F,\bar \F) \in \left\{ 
\begin{array}{cc}
{\rm SO}^*(2n)~, &~~\e=+1\,,
\\
{\rm Sp}(n,{\mathbb R})~,~& ~~\e=-1\,. 
\end{array}
\right.
\label{nc-cosetrep4}
\eea

\subsection{The symmetric  space 
${\rm SO}_0(n,2)/{\rm SO}(n) \times {\rm SO}(2)$}
\label{noncompactquadric1}

As a real manifold, the Hermitian symmetric space\footnote{Here
${\rm SO}_0(n,2)$ denotes the connected component of the
identity in ${\rm O}(n,2)$.} 
${\rm SO}_0(n,2)/{\rm SO}(n) \times {\rm SO}(2)$ can be identified
\cite{Hua1} with an open subset of the oriented Grassmann manifold 
$\tilde{G}_{2,n+2} (\mathbb R )$.  In the notation of subsection \ref{1.3},
see eqs. (\ref{x}) and (\ref{er-quadric1-com}), consider 
the following domain in $\tilde{G}_{2,n+2} (\mathbb R )$:
\bea
\cM=\Big\{ x \in \tilde{G}_{2,n+2} (\mathbb R )~, \qquad 
x^{\rm T}\, \O \,x >0 \Big\} ~,
\qquad 
\O =   \left(
\begin{array}{cc}
- \mathbbm{1} _n & 0\\
0 & \mathbbm{1} _2
\end{array}
\right)~.
\label{M}
\eea
By construction, this domain is a transformation space
of the group ${\rm O}(n,2)$.
It follows from (\ref{M}) that the $2\times 2$ block $\hat{x}$ 
is non-singular,
$\hat{x} \in {\rm GL}(2, {\mathbb R})$.
As a topological space, 
the domain $\cM$ consists of two connected 
components with empty intersection,
$\cM =\cM^+ \cup \cM^-$, defined as
\bea
\cM^+=\Big\{ x \in \cM~, \quad \det \hat{x} > 0\Big\}~;
\qquad 
\cM^-=\Big\{ x \in \cM~, \quad \det \hat{x} < 0\Big\}~. 
\eea
With respect to the action of the subgroup ${\rm SO}_0(n,2) \in {\rm O}(n,2)$, 
the sub-domains $\cM^+$ and $\cM^-$ can be shown to be  the orbits.
We identify the Hermitian symmetric space
${\rm SO}_0(n,2)/{\rm SO}(n) \times {\rm SO}(2)$ 
 with the orbit $\cM^+$.

${}$For $x \in \cM^+$, its  equivalence class  contains a
matrix 
\bea
u= (u^I{}_\a) =
\left(
\begin{array}{c}
\tilde{u} \\
\hat{u} 
\end{array}
\right)~ 
\eea
constrained to obey
\bea
u^{\rm T}\, \O \,u
=\hat{u}^{\rm T} \hat{u} - \tilde{u}^{\rm T} \tilde{u} 
= \mathbbm{1} _2~, \qquad 
\det \hat{u} > 0~.
\label{quadric-constr1}
\eea
We can further represent 
\bea
\hat{u} =s\,h~, \qquad s=s^{\rm T}= (s_{\a \b}) ~, 
\qquad h \in {\rm SO}(2)~,
\label{pd-o}
\eea
with $s$ positive definite.
Under eq. (\ref{quadric-constr1}), the `gauge'
freedom  (\ref{er-quadric1-com}),
reduces to 
\bea
u~ \sim~ u g ~, \qquad g \in {\rm SO}(2)~. 
\label{er-quadric2}
\eea
This residual freedom can be completely fixed by choosing 
in (\ref{pd-o}) $h=\mathbbm{1} _2$.

Let us introduce the coset representative 
\bea
G(u) = \left(
\begin{array}{cc}
\mathbbm{1}_n +  \tilde{u} h^{-1}
\l(s) h \tilde{u}^{\rm T} ~  ~& ~\tilde{u} h^{-1}\\
h \,\tilde{u}{}^{\rm T} ~~& ~s 
\end{array}
\right)~, \qquad
\l (s) = \frac{\mathbbm{1} _2}{\mathbbm{1}_2 +s}~.
\label{quadric-cosetrep}
\eea
The crucial  property of $G(u)$ is
\bea
G(u)\, u_0 =
 \left(
\begin{array}{c}
\tilde{u} h^{-1}\\
s 
\end{array}
\right)~, \qquad
u_0
= \left(
\begin{array}{c}
\tilde{u}_0 \\
\hat{u}_0 
\end{array}
\right)=\left(
\begin{array}{c}
0 \\
\mathbbm{1} _2
\end{array}
\right)~.
\eea

Introduce a complex $(n+2)$-vector 
\bea
w=(w^I) := (u^I{}_1 +{\rm i}\, u^I{}_2)
= \left(
\begin{array}{c}
\tilde{w} \\
\hat{w} 
\end{array}
\right)\equiv \left(
\begin{array}{c}
\vf \\
z 
\end{array}
\right)~, \qquad \vf = (\vf^i)~, 
\quad z = \left(
\begin{array}{c}
z_1 \\
z_2
\end{array}
\right)~.
\eea
Now, the first equation in (\ref{quadric-constr1}) is equivalent to
\bea
w^{\rm T} \O \,w &=& -\vf^{\rm T} \vf +(z_1)^2 +(z_2)^2 =0~, 
\label{quad1}
\\
w^\dagger \O \,w& =& -\vf^\dagger \vf +|z_1|^2 +|z_2|^2 = 2~.
\label{quad2}
\eea
The `gauge' freedom (\ref{er-quadric2}) becomes 
\be
w ~\sim ~{\rm e}^{{\rm i}\, \s}w~, \qquad \s \in \mathbb R ~.
\label{er-quadric3}
\ee

In what follows, we choose the gauge condition 
$h=\mathbbm{1} _2$.
Recall that  $s =s^{\rm T}$ is positive definite, that is 
\bea
s = \left(
\begin{array}{cc}
s_{11}& s_{12} \\
s_{12} & s_{22}
\end{array}
\right)~, \qquad s_{11} >0~, \quad s_{22} >0~,
\quad s_{11}s_{22} - (s_{12})^2 >0~.
\eea
These results tell us that both the components of $z$,
\bea
 \left(
\begin{array}{c}
z_1 \\
z_2
\end{array}
\right)
:= \left(
\begin{array}{c}
(s\,h)_{11} +{\rm i} \,(s\,h)_{12} \\
(s\,h)_{12}  +{\rm i} \,(s\,h)_{22} 
\end{array}
\right)
= \left(
\begin{array}{c}
s_{11} +{\rm i} \,s_{12} \\
s_{12}  +{\rm i} \,s_{22} 
\end{array}
\right)~,
\eea
are non-vanishing, $z_{1,2} \neq 0$. If we further introduce new variables
\bea
z_\pm = z_1 \pm {\rm i} \,z_2~,
\eea
then one readily sees\footnote{In deriving eq. (\ref{quad3}), 
we have fixed the gauge freedom  (\ref{er-quadric3})
by imposing the condition $h=\mathbbm{1} _2$. 
In general, the expression for $z_-$ is as follows:
$z_- = {\rm e}^{{\rm i}\,\s}(s_{11} +s_{22} )$, 
with $\s$ a real parameter.}
\bea
z_- =\overline{z_-}=s_{11} +s_{22} >0~, \qquad
\Big| \frac{z_+}{z_-}\Big|^2 <1~.
\label{quad3}
\eea
${}$Finally, introducing projective variables 
\bea
\F = \frac{\vf}{z_-}~, \qquad \r =\frac{z_+}{z_-}~,
\eea
the equations (\ref{quad1}) and (\ref{quad2}) turn into
\bea
-\F^{\rm T}\F +\r&=&0~, \\
-2 \F^\dagger \F  +1 +|\r|^2 &=&\frac{4}{(z_-)^2}~.
\label{z-}
\eea
In conjunction with eq. (\ref{quad3}), we now  
see that the $n$ complex variables $\F =(\F^i)$ span the domain
\bea
-2 \F^\dagger \F  +1 +|\F^{\rm T}\F|^2 >0~, \qquad 
|\F^{\rm T}\F|<1~.
\eea
These conditions define a classical bounded symmetric 
domain \cite{Cartan,Hua1,Hua2}.

Let us consider an isometry transformation $g \in {\rm SO}_0(n,2)$.
\bea
g= \left(
\begin{array}{cc}
A & B \\
C& D
\end{array}
\right) ~, \qquad 
g^{\rm T} \O \,g = \O~,
\label{isom-non-com1}
\eea
or equivalently 
\bea
A^{\rm T} A- C^{\rm T} C = \mathbbm{1}_n~, 
\qquad 
D^{\rm T} D- B^{\rm T} B = \mathbbm{1}_2~, 
\qquad A^{\rm T} B= C^{\rm T} D ~. 
\label{isom-non-com2}
\eea 
The fact that $g$ belongs to the connected component of the identity
in  ${\rm O}(n,2)$, is expressed as follows:
\be
\det A >0~, \qquad \det D >0~.
\ee
The linear action $u \to u'=g u$ induces
the holomorphic fractional linear transformation
\bea
\F ~\to~\F' &=& \Big\{ (1,-{\rm i}) \Big( C\, \F + D \,\G(\F)
\Big) \Big\}^{-1}  
\Big\{ A \,\F + B\,\G(\F) \Big\} ~,
\non \\
\G (\F) &=& \hf \left(
\begin{array}{c}
1 +\F^{\rm T}\F \\
{\rm i} (1- \F^{\rm T}\F) 
\end{array}
\right)~.
\label{IVholom}
\eea
Here we have used the fact that $z_-$ transforms as follows:
\be
\frac{z'_-}{z_-} = (1,-{\rm i}) \Big( C\, \F + D \,\G(\F)
\Big) \equiv {\rm e}^{\L(\F)}~.
\label{z--transfo}
\ee
Unlike $z_-$, its transform $z'_-$ is no longer real, 
but its phase is a gauge degree of freedom.

The K\"ahler potential  \cite{Hua1,Hua2,Perelomov} is 
\be
K(\F, \bar \F ) =-\hf \ln \Big(1-2 \F^\dagger \F  +|\F^{\rm T}\F|^2 \Big)~.\label{kahler-nonq}
\ee
Under the holomorphic isometry transformation (\ref{IVholom}), 
it changes as 
\bea 
K(\F', \bar \F' ) =K(\F, \bar \F ) + \L(\F) + {\bar \L}(\bar \F )~, 
\eea
with $\L(\F) $ given in (\ref{z--transfo}).
This can be seen from the identity
\bea
1-2 \F'^\dagger \F'   +|\r'|^2 
=\Big(1 -2 \F^\dagger \F   +|\r|^2 \Big)\,
\Big| \frac{z_-}{z'_-} \Big|^2~,
\label{structural-non-cq}
\eea
in conjunction with eq. (\ref{z--transfo}).

Let us turn to the problem of expressing 
the coset representative (\ref{quadric-cosetrep}) 
in terms of the complex coordinates introduced above. 
In the gauge $h=\mathbbm{1} _2$
we can rewrite $G(u)$ as 
\bea
G(\F, \bar \F) = \left(
\begin{array}{cc}
\underline{s} ~  ~& ~\tilde{u} 
\\
\tilde{u}{}^{\rm T} ~~& ~s 
\end{array}
\right)
\equiv
 \left(
\begin{array}{cc}
\cA ~  ~& ~\cB
\\
\cC ~~& ~ \cD
\end{array}
\right)
~, 
\label{quadric-cosetrep2}
\eea
where 
\be
\underline{s}^2 = \mathbbm{1} _n +\tilde{u}\, \tilde{u}{}^{\rm T} ~, 
\qquad 
s^2 = \mathbbm{1} _2 + \tilde{u}{}^{\rm T}\tilde{u}~.
\ee
${}$For the matrix blocks in (\ref{quadric-cosetrep2}) we then get
\bea
\cA &=& \sqrt{
\mathbbm{1} _n + \frac{ z_-^2}{2} (\F\F^\dagger + {\bar \F} \F^{\rm T})
}~, \non \\
\cB&=& \hf z_- (\F\,, \,{\bar \F})\,\g~,
\qquad \cC = \hf z_-  \g^\dagger 
 \left(
\begin{array}{c}
\F^\dagger
\\
\F^{\rm T}
\end{array}
\right)
~,
\non\\
\cD &=&\hf \g^\dagger \sqrt{
\mathbbm{1} _2 +  \frac{ z_-^2}{2} \,\D
}\, \g~, 
\label{matrix1}
\eea
where
\bea
\g=  \left(
\begin{array}{cr}
1~  & ~-{\rm i}
\\
1 ~& ~ {\rm i}
\end{array}
\right)~, \qquad 
\D = \left(
\begin{array}{cc}
\F^\dagger \F  ~& ~\overline{\F^{\rm T}\F}  \\
\F^{\rm T}\F~& ~ \F^\dagger \F
\end{array}
\right)~.
\eea
Eq. (\ref{z-}) gives the expression for $z_-$ in terms of $\F$ and its 
conjugate. The isometry transformation $G(\F, \bar \F) \in {\rm SO}_0(n,2)$
maps the origin, $\F_0=0$, to the point $\F$. On a generic point 
$\U$ of the symmetric space, it acts by the rule:
\bea
\U ~\to~\U' &=& \Big\{ (1,-{\rm i}) \Big( \cC\, \U + \cD \,\G(\U)
\Big) \Big\}^{-1}  
\Big\{ \cA \,\U + \cB\,\G(\U) \Big\} ~, 
\label{trans-nonq}
\eea
with the two-vector $\G(\U)$ defined similarly to (\ref{IVholom}).

It is easy to check that the matrix $\cD$ in (\ref{matrix1}) possesses 
the following equivalent representation:
\bea
\cD = \frac{1}{4} z_- \g^\dagger F \, \g ~, \qquad 
F  = \left(
\begin{array}{cc}
1 ~& ~\overline{\F^{\rm T}\F}  \\
\F^{\rm T}\F~& ~ 1
\end{array}
\right)~.
\label{matrix2}
\eea
One can also directly verify that 
\bea
 \left(
\begin{array}{c}
\F^\dagger  \\
\F^{\rm T}
\end{array}
\right)\, \cA = \hf \g \, \cD\, \g^\dagger 
 \left(
\begin{array}{c}
\F^\dagger  \\
\F^{\rm T}
\end{array}\right) 
= \hf z_- F 
\left(
\begin{array}{c}
\F^\dagger  \\
\F^{\rm T}
\end{array}
\right)~.
\label{matrix3}
\eea
Another useful piece of  information is that the first equation 
in (\ref{isom-non-com2}) is equivalent, in the case of $G(\F, \bar \F)$, to 
\bea
\cA^{\rm T} \cA = \mathbbm{1} _n +\hf z_-^2 
\Big(\F , \,{\bar \F} \Big) 
\left(
\begin{array}{c}
\F^\dagger  \\
\F^{\rm T}
\end{array}
\right)~.
\label{matrix4}
\eea

\sect{The (co)tangent bundle over ${\rm U}(n,m)/{\rm U}(n)\times {\rm U}(m)$}
The K{\"a}hler potential in this cases is given as
\begin{eqnarray}
 K(\Phi,\Phi^\dagger)=-\ln\det(\mathbbm{1} _m-\Phi^\dagger\Phi)
 =-\ln\det(\mathbbm{1}_n - \Phi\Phi^\dagger)~, 
\label{kahler-nonu}
\end{eqnarray}
where $\Phi=(\Phi^{i\alpha})$
and  $\Phi^\dagger=(\bar{\Phi}^{\bar{\alpha}\bar{i}})$.
The K{\"a}hler metric can be read off to be 
\begin{eqnarray}
 g_{i\alpha,\bar{\beta}\bar{j}}
 =\left({\mathbbm{1} _m \over 
 {\mathbbm{1}}_m-\Phi^\dagger\Phi}\right)_{\alpha\bar{\beta}}
\left({\mathbbm{1} _n \over \mathbbm{1} _n-\Phi\Phi^\dagger}\right)_{\bar{j}i}\,.
 \label{metric-nonu}
\end{eqnarray}
where we have used identities similar to eq. (\ref{identity1}),
\be
\frac{\mathbbm{1} _n}{  \mathbbm{1} _n - \F \F^\dagger }\,\F 
= \F \,\frac{\mathbbm{1} _m}{  \mathbbm{1} _m - \F^\dagger \F}~,
\qquad 
\F^\dagger \,\frac{\mathbbm{1} _n}{  \mathbbm{1} _n - \F \F^\dagger }
= \frac{\mathbbm{1} _m}{  \mathbbm{1} _m - \F^\dagger \F}\,\F^\dagger~. \label{identity2}
\ee
The explicit structure of the K\"ahler potential forces us to choose
\begin{eqnarray}
K(\Upsilon,\breve{\Upsilon})
 =
- \ln \det ( \mathbbm{1} _m - \breve{\U}^{\rm T} \U) 
= -\ln \det ( \mathbbm{1} _n - \U \breve{\U}^{\rm T} )
 \label{k-nonu}
\end{eqnarray}
in the action (\ref{action}).
The equations of motion for the auxiliary superfields are: 
\begin{eqnarray}
 \oint 
 \frac{{\rm d}\zeta }{ \zeta}\zeta^n 
 \left({\bf
  1}_m - \breve{\Upsilon}_*^{\rm T}\Upsilon_*
 \right)^{-1}\breve{\Upsilon}_*^{\rm T}=0~, \qquad n\ge 2~. 
 \label{eq-mot:nonu}
\end{eqnarray}
Their solution, $\U_*(\z)$, is obtained by 
applying 
the coset representative (\ref{nc-cosetrep2}) to (\ref{u-0}). 
\begin{eqnarray}
 \Upsilon_*=\left\{(\mathbbm{1} _n+\Phi\lambda s^2
\Phi^\dagger)\S_0 s^{-1}\z + \Phi \right\}
 (\mathbbm{1} _m + s\Phi^\dagger \S_0 s^{-1}\zeta)^{-1}\,.
\label{sol-nonu}
\end{eqnarray}
${}$From here one reads off the
tangent vector at $\F$
\begin{eqnarray}
 \Sigma \equiv  \frac{\partial \Upsilon_* }{ \partial \zeta}{\Bigg
  |}_{\zeta=0}
 =(\mathbbm{1} _n + \Phi\lambda s^2 \Phi^\dagger - \Phi s \Phi^\dagger) \Sigma_0 s^{-1}
&=&(\mathbbm{1} _n - \Phi\lambda s \Phi^\dagger ) \Sigma_0 \,s^{-1} \non \\
 &=& \underline{s}^{-1} \Sigma_0 \,s^{-1} ~.
\label{Sigma-nonu}
\end{eqnarray}
This result allows us to express $\S_0$ in terms of $\S$. 
Let us substitute the solution (\ref{sol-nonu}) into the potential
(\ref{k-nonu}).
\begin{eqnarray}
 K (\Upsilon_*,\breve{\Upsilon}_*)
 &=&-\ln \det \Big(\mathbbm{1} _m -\breve{\Upsilon}_*^{\rm T} \Upsilon_* \Big) \nonumber \\
&=&-\ln \det \Big(\mathbbm{1} _m  - \Phi^\dagger \Phi + s^{-1}\Sigma_0^\dagger
\Sigma_0 s^{-1}\Big) 
\nonumber \\
&& +\ln \det \Big(\mathbbm{1} _m - {1 \over \zeta}s^{-1}\Sigma_0^\dagger \Phi s
\Big) + \ln\det  \Big(\mathbbm{1} _m + s\Phi^\dagger \Sigma_0
s^{-1}\zeta \Big)
\,. 
\label{k-nonu2}
\end{eqnarray}
Here we have used eqs. (\ref{nc-cosetrep}) and (\ref{nc-s}), and their corollary
\be
-\F s^2 \F^\dagger +(\mathbbm{1} _n + \Phi\lambda s^2 \Phi^\dagger )^2 =\mathbbm{1} _n\,. 
\ee

${}$From (\ref{k-nonu2}) we obtain the tangent bundle action
\bea
 S&=& \int {\rm d}^8z \left\{
 K(\F , \F^\dagger) 
 - \ln \det \Big( \mathbbm{1} _m
 +(\mathbbm{1} _m -\Phi^\dagger \Phi)^{-1} \Sigma^\dagger 
  (\mathbbm{1} _n - \F \Phi^\dagger )^{-1}\Sigma \Big) \right\} \non \\
&=& \int {\rm d}^8z \left\{
 K(\F , \F^\dagger) 
- \ln \det \Big( \mathbbm{1} _n+ (\mathbbm{1} _n - \F \Phi^\dagger )^{-1}
\Sigma \,
(\mathbbm{1} _m -\Phi^\dagger \Phi)^{-1} \Sigma^\dagger 
\Big) \right\}~, ~~~~~
\label{k-nonu3}
\eea
where $K(\Phi,\Phi^\dagger)$ is the K{\"a}hler potential, eq. (\ref{kahler-nonu}).
Unlike the compact case, eq. (\ref{Grassmann-global}),
no restrictions on the tangent variables $\S$ occur.

By construction, the theory with action (\ref{k-nonu3}) is invariant under
the isometry group U$(n,m)$ of the base manifold. However, the symmetry 
properties are somewhat hidden in the  action constructed.
To make them manifest, it is useful to decompose the tangent vectors with respect 
to the vielbein (\ref{vielbein-non}):
\bea
\S ~\to ~ \tilde{\S} = \underline{s} \,\S \,s~, \qquad 
\S^\dagger ~\to ~\tilde{\S}^\dagger = s\, \S^\dagger \,\underline{s}~,
\label{tildeSigma-non}
\eea  
similarly to the compact case. Given an isometry transformation, 
it proves to act on $\tilde \S$ as an induced local transformation
from the isotropy group ${\rm U}(n)\times {\rm U}(m)$.  Such a transformation 
acts on $\tilde \S$ as follows:
\bea
  \tilde{\S} ~\to ~ g_L \tilde{\S} \,g_R, \qquad \quad
  g_L \,\in {\rm U}(n)~, \qquad   g_R \in {\rm U}(m)~.
\eea
This implies that the  tangent bundle action 
\bea
 S&=& \int {\rm d}^8z \left\{
 K(\F , \F^\dagger) 
- \tr \ln  \Big( \mathbbm{1} _m
+ \tilde{\S}^\dagger \tilde{\S}
\Big)
 \right\} 
\non \\ 
&=&
 \int {\rm d}^8z \left\{
 K(\F , \F^\dagger) 
-  \tr \ln  \Big( \mathbbm{1} _n
+ \tilde{\S} \tilde{\S}^\dagger 
\Big) \right\}  
\label{action-gr-2-non}
\eea
is indeed U$(n,m)$-invariant.
In appendix A, the curvature tensor of 
the symmetric space ${\rm U}(n,m)/{\rm U}(n)\times {\rm U}(m)$ is computed,
eq. (\ref{curvature-non-comp}). 
It follows from (\ref{curvature-non-comp})
that the Taylor expansion of (\ref{action-gr-2-non}) in powers 
of $\S$ and $\bar \S$
(in the domain $\tilde{\S}^\dagger \tilde{\S} < \mathbbm{1}_m$) 
 can be represented in  the universal form
(\ref{act-tab}).

Derivation of the cotangent bundle formulation is very similar 
to the compact case. One introduces the first-order action 
\begin{eqnarray}
 S&=& \int {\rm d}^8z \left\{
 K(\F , \F^\dagger) 
- \tr \ln  \Big( \mathbbm{1} _m + 
 s  U^\dagger \underline{s}^2 U s \Big) +\hf \tr \,(U \j) 
 +\hf \tr \,(\j^\dagger U^\dagger  ) \right\}~,~~~
 \label{cot-gr0-non}
\end{eqnarray}
where $U=(U^{i\alpha})$ 
 is a complex unconstrained superfield, and 
$\j=(\psi_{\alpha i})$ is a chiral superfield. 
By  construction, $U$ is a tangent vector at the point $\Phi$ of
 the base manifold.
Thus $\psi$ should be a one-form at the same point.
In order to derive the cotangent bundle formulation, 
the unconstrained superfield variables, $U$ and $U^\dagger$,
 have to be eliminated with the aid 
of their equations of motion. This procedure is 
considerably simplified if one deals with (co)tangent vectors 
decomposed 
with respect to the vielbein  (\ref{vielbein-non}), 
\bea
U ~\to ~ \tilde{U} = \underline{s} \,U \,s~, \qquad
\j ~\to ~ \tilde{\j} =s^{-1} \j\, \underline{s}^{-1} ~.
\eea  

Repeating the technical steps described in the case 
of the Grassmannian, we end up with the cotangent bundle 
action
\begin{eqnarray}
 S&=& \int {\rm d}^8z \left\{
 K(\F , \F^\dagger) 
+ \tr \ln  \Big( \mathbbm{1} _n 
+  \sqrt{\mathbbm{1}_n -\tilde{\j}^\dagger \tilde{\j}} \Big)
-\tr \,\sqrt{\mathbbm{1}_n -\tilde{\j}^\dagger \tilde{\j}} 
 \right\} \non \\
 &=& \int {\rm d}^8z \left\{
 K(\F , \F^\dagger) 
+ \tr \ln  \Big( \mathbbm{1} _m 
+  \sqrt{\mathbbm{1}_m - \tilde{\j}\tilde{\j}^\dagger  } \Big)
-\tr \,\sqrt{\mathbbm{1}_m - \tilde{\j} \tilde{\j}^\dagger  } 
 \right\} ~.
 \label{cot-gr0-3-non}
\end{eqnarray}
This action is well-defined under the following 
constraints:
\bea
\F^\dagger \F < \mathbbm{1}_m~, \qquad
\tilde{\j}^\dagger \tilde{\j} < \mathbbm{1}_n~.
\eea
${}$From here it follows that the hyperk\"ahler structure is
defined on a unit ball of the zero section of the cotangent bundle
over ${\rm U}(n,m)/{\rm U}(n)\times {\rm U}(m)$.

Consider the simplest case, $n=m=1$, analysed in \cite{K3}. 
This choice corresponds to the 
complex hyperbolic line $\bH^1={\rm SU}(1,1)/{\rm U}(1)$.
It is known that any compact Riemann surface $\bm \S$  of genus $g>1$ can be obtained 
from $\bH^1$ by factorization  with respect to some discrete 
subgroups of ${\rm SU}(1,1)$, see e.g. \cite{FK}. 
Using the hyperk\"ahler metric constructed on the open disc bundle in 
 $T^*\bH^1$, we then can generate a hyperk\"ahler 
 structure defined on  an open neighbourhood of the zero section 
 of the cotangent bundle $T^*{\bm \S}$.

\sect{The (co)tangent bundle over ${\rm SO}^*(2n)/{\rm U}(n)$ and 
${\rm Sp}(n,{\mathbb R})/{\rm U}(n)$}

${}$For the non-compact Hermitian symmetric spaces
${\rm SO}^*(2n)/{\rm U}(n)$ and ${\rm Sp}(n,{\mathbb R})/{\rm U}(n)$,
the K{\"a}hler potentials are 
\begin{eqnarray}
 K(\Phi,\Phi^\dagger)=-\ln\det(\mathbbm{1} _n-\Phi^\dagger\Phi)=-\ln\det(\mathbbm{1}_n-\Phi\Phi^\dagger)\, ,
\label{kahler-nonso}
\end{eqnarray}
with $\F=(\F^{ij})$ a complex $n\times n$ matrix 
obeying the constraint (\ref{nc-sa}).
The K{\"a}hler metric is readily obtained to be 
\begin{eqnarray}
 g_{ik,\bar{l}\bar{j}}=\left({\mathbbm{1} _n \over {\bf
1}_n-\Phi^\dagger\Phi}\right)_{k\bar{l}}
\left({\mathbbm{1} _n \over \mathbbm{1} _n-\Phi\Phi^\dagger}\right)_{\bar{j}i}\,,
\end{eqnarray}
where we have used eq. (\ref{identity2}).

In complete analogy with the compact case, 
the (co)tangent bundle formulations can be deduced from 
those already derived for the symmetric space
${\rm U}(n,m)/{\rm U}(n)\times {\rm U}(m)$,  
eqs. (\ref{k-nonu3}) and (\ref{cot-gr0-3-non}). 
In these actions, one should simply set $m=n$ and 
require the tangent linear $\S$ and cotangent chiral $\j$ 
variables to obey the algebraic constrains
\be
\S^{\rm T} =- \e \,\S~, \qquad 
\j^{\rm T} = - \e \, \j~.
\ee
The cotangent variables $\tilde \j$ in (\ref{cot-gr0-3-non})
can now be  expressed via $\j$ in two equivalent forms:
\be
\tilde{\j} =s^{-1} \j\, \underline{s}^{-1} = s^{-1} \j\, \bar{s}^{-1} ~.
\ee
One can readily fill  the missing detail.

\sect{The (co)tangent bundle over 
${\rm SO}_0(n,2)/{\rm SO}(n) \times {\rm SO}(2)$}
\label{4}
In accordance with the discussion in section
\ref{noncompactquadric1},
the K\"ahler potential   \cite{Hua1,Hua2,Perelomov}  is 
\be
K(\F, \bar \F ) =-\hf \ln \frac{4}{z_-^2}~, \qquad 
\frac{4}{z_-^2}=
1-2 \F^\dagger \F  +|\F^{\rm T}\F|^2 
~.
\label{kahler-nonq2}
\ee
Its ${\cal N}=2$ extension is given by
\begin{eqnarray}
K(\Upsilon,\breve{\Upsilon})=
-\hf \ln \left(1 - 2 \breve{\Upsilon}^{\rm T} \Upsilon
+\Upsilon^{\rm T} \Upsilon
\,\breve{\U}^{\rm T} \breve{\U} \right)~. 
\label{kahler-nonq3}
\end{eqnarray} 

The equations of motion for the auxiliary superfields are 
\begin{eqnarray}
0=
\oint {{\rm d}\zeta \over \zeta}\zeta^n
 \frac{ -\breve{\Upsilon}
 + \Upsilon
 \,\breve{\U}^{\rm T} \breve{\U}}
 {  1 - 2 \breve{\Upsilon}^{\rm T} \Upsilon
+\Upsilon^{\rm T} \Upsilon
\,\breve{\U}^{\rm T} \breve{\U} }~,
~~~~n\ge 2\,. 
\label{auxiliary-nonq}
\end{eqnarray}
Their solution, $\U_*(\z)$,  is obtained from (\ref{trans-nonq})
by replacing $\U \to \z \S_0$ and $\U' \to \U_*(\z)$.
This gives
\bea
\U_*(\z) = \frac{ \F +(2/ z_-)\,   
{\cal A} \Sigma_0 \,\zeta + 
\bar{\Phi} \,\Sigma_0^{\rm T}\Sigma_0\, \zeta^2
}
{1 +2 \F^\dagger \S_0\, \z + 
\overline{\F^{\rm T}\F}\, \S_0^{\rm T} \S_0 \, \z^2 }~.
\label{xy}
\eea
${}$From here 
\begin{eqnarray}
\Sigma=\frac{ {\rm d} \Upsilon_*}{{\rm d} \zeta}{\Big |}_{\zeta=0}={2 \over z_-}\,
{\cal A}\Sigma_0 - 2\Phi (\Phi^{\dagger}\Sigma_0)~.
\end{eqnarray}
Now, one can express $\S_0$ in (\ref{xy}) via $\S$ 
by making use of eqs. (\ref{matrix3}) and (\ref{matrix4}). 
One thus derives 
\bea
\F^\dagger \S_0 &=& \frac{1}{4}z_-^2 \Big\{ 
\F^\dagger \S - \overline{\F^{\rm T}\F}\,\F^{\rm T}\S \Big\}~, \non \\
\F^{\rm T} \S_0 &=& \frac{1}{4}z_-^2 \Big\{ (1-2|\F|^2 )\,\F^{\rm T} \S
+ \F^{\rm T}\F\,\F^\dagger \S \Big\}~, \non \\
\S_0^{\rm T} \S_0 &=& \frac{1}{4}z_-^2 \S^{\rm T} \S~, 
\label{S-0-->S} \\
\S_0^{\dagger} \S_0 &=& \frac{1}{4}z_-^2 \S^{\dagger} \S  
- \frac{1}{8}z_-^4
\Big\{ (1-2|\F|^2 )\,\F^{\rm T} \S \, \S^\dagger {\bar \F} \non \\
&& \qquad \qquad 
+\overline{\F^{\rm T}\F}\,\F^{\rm T}\S \, \S^\dagger \F 
+ \F^{\rm T}\F\,\F^\dagger \S \, \S^\dagger {\bar \F} 
- \F^\dagger \S \, \S^\dagger \F \Big\} ~. \non
\eea
The final form for $\U_* $ is as follows:
\bea
\U_*(\z) = \frac{ 
\F +  \zeta \S + (z_-^2/2) \Big\{ \z \F(  \F^\dagger \S 
- \overline{\F^{\rm T}\F}\,\F^{\rm T}\S ) 
+\hf \z^2 {\bar \F}\, \S^{\rm T} \S \Big\}
}
{1 + \hf z_-^2  \Big\{ \z ( \F^\dagger \S 
- \overline{\F^{\rm T}\F}\,\F^{\rm T}\S ) 
+\hf \z^2 \overline{\F^{\rm T}\F} \S^{\rm T} \S \Big\} 
}~.
\label{xy2}
\eea

We now have to evaluate the superfield Lagrangian 
\bea
L= \frac{1}{2\p {\rm i}}
 \oint \frac{ {\rm d} \zeta }{ \zeta} K(\U_*,\breve{\U}_*)~.
\eea
Considerations similar to those used to derive eq.
(\ref{structural-non-cq}), can  be used to show that
\bea
K(\U_*,\breve{\U}_*)&=& K(\F , \bar \F) 
-\hf \ln \Big( 1 +2 \S_0^\dagger \S_0 +|\S^{\rm T}_0\S_0|^2 \Big)
+\hf \ln x(\z) +\hf \ln \breve{x} (\z)~, ~~~~\\
x(\z) &=& 1 +2 \F^\dagger \S_0\, \z + 
\overline{\F^{\rm T}\F}\, \S_0^{\rm T} \S_0 \, \z^2 ~.
\non
\eea
Clearly, the last two terms in the expression for $K(\U_*,\breve{\U}_*)$
do not contribute to $L$.
With the aid of the third and fourth relations in (\ref{S-0-->S}) we obtain
\bea
L & = &  K(\F , \bar \F) 
-\hf \ln \Big( 1 + 
 \frac{1}{2}z_-^2 \S^{\dagger} \S  
- \frac{1}{4}z_-^4
\Big\{ (1-2|\F|^2 )\,\F^{\rm T} \S \, \S^\dagger {\bar \F} \non \\
&& 
+\overline{\F^{\rm T}\F}\,\F^{\rm T}\S \, \S^\dagger \F 
+ \F^{\rm T}\F\,\F^\dagger \S \, \S^\dagger {\bar \F} 
- \F^\dagger \S \, \S^\dagger \F
+\frac{1}{16}z_-^4 |\S^{\rm T} \S|^2
 \Big\} \Big)~.
\eea
This can be transformed to a more geometric form 
 by taking two observations into account. 
${}$First, using the expression for the metric 
of the non-compact quadric surface
\begin{eqnarray}
 g_{i\bar{j}}=
   \frac{1}{4}z_-^2
 \delta_{i\bar{j}}
 -  \frac{1}{8}z_-^4
 \left\{
 \Phi^i\bar{\Phi}^{\bar{j}} (1-2|\F|^2) 
- \bar{\Phi}^{\bar{i}} \F^j 
+ \Phi^i\Phi^{j}(\overline{\Phi^{\rm T}\Phi})
 +\bar{\Phi}^{\bar{i}}\bar{\Phi}^{\bar{j}}(\Phi^{\rm T}\Phi) 
\right\}\,,
\end{eqnarray}
we find
\begin{eqnarray}
g_{i\bar{j}}\Sigma^i\bar{\Sigma}^{\bar{j}}=
  \frac{1}{4}z_-^2
 |\Sigma|^2
&-& 
 \frac{1}{8}z_-^4
\Big\{
  (1-2|\F|^2) 
 |\Phi^{\rm T}\Sigma|^2
 -|\Phi^\dagger \Sigma|^2 
\nonumber \\
 &+& (\Phi^{\rm  T}\Phi)
 (\Phi^\dagger\bar{\Sigma})(\Phi^\dagger\Sigma)
   +  (\overline{\Phi^{\rm T}\Phi}) 
 (\Phi^{\rm T}\Sigma)(\Phi^{\rm T}\bar{\Sigma})
  \Big\}~.
\end{eqnarray}
Thus 
\bea
L & = &  K(\F , \bar \F) 
-\hf \ln \Big( 1 + 2 g_{i\bar{j}}\Sigma^i\bar{\Sigma}^{\bar{j}}
+ \frac{1}{16}z_-^4
|\S^{\rm T} \S|^2 \Big)~.
\label{L}
\eea
Second, we note that $z_-^4|\S^{\rm T} \S|^2 $ must be  a scalar 
field on the tangent bundle, and therefore it can be expressed solely in terms 
of the tensor quantities: the holomorphic tangent vector $\S^i$ and 
its conjugate, the Riemann metric $g_{i\bar{j}}$, and finally the Riemann 
curvature
$R_{i\bar{j}k\bar{l}}\equiv
\partial_k\bar{\partial}_{\bar{l}}g_{i\bar{j}}
 -g^{m\bar{n}}\partial_{m}g_{i\bar{j}}\bar{\partial}_{\bar{n}}g_{k\bar{l}}$, 
 with $\partial_i=\partial / \partial \Phi^i$.
It is sufficient to determine such an expression
at any given point of the base manifold, say at $\F=0$, 
since the base manifold is a symmetric space.
This gives 
\begin{eqnarray}
2(g_{i\bar{j}}\Sigma^i\bar{\Sigma}^{\bar{j}})^2 -\hf R_{i\bar{j}k\bar{l}}
 \Sigma^i\bar{\Sigma}^{\bar{j}}\Sigma^k\bar{\Sigma}^{\bar{l}}
&=&
 \frac{1}{16}z_-^4
|\S^{\rm T} \S|^2 ~.
\end{eqnarray}

As a result, we obtain the tangent bundle action
\begin{eqnarray}
 S&=&\int {\rm d}^8z \left\{
 K(\Phi,\bar{\Phi})
- \hf \ln \Big( 1+ 2g_{i\bar{j}}\Sigma^i\bar{\Sigma}^{\bar{j}}
+2 (g_{i\bar{j}}\Sigma^i\bar{\Sigma}^{\bar{j}})^2
-\hf R_{i\bar{j}k\bar{l}}
 \Sigma^i\bar{\Sigma}^{\bar{j}}\Sigma^k\bar{\Sigma}^{\bar{l}} \Big)
\right\}\,, ~~~~ 
\label{tan-a-nonq}
\end{eqnarray}
where $K(\Phi,\bar{\Phi})$ is the K{\"a}hler potential of the 
non-compact quadric
 surface, eq.  (\ref{kahler-nonq2}).
It follows from (\ref{L}) that the action is well-defined on the  tangent bundle.

${}$Finally,  let us dualize the tangent bundle action (\ref{tan-a-nonq}). 
In order to do that we replace the action (\ref{tan-a-nonq}) with
\begin{eqnarray}
S=\int {\rm d}^8z {\bigg \{}
 K(\Phi,\bar{\Phi})
 &-& \hf \ln \bigg( 1+ 2g_{i\bar{j}}U^i\bar{U}^{\bar{j}}
 +2 (g_{i\bar{j}}U^i\bar{U}^{\bar{j}})^2
 -\hf R_{i\bar{j}k\bar{l}} 
 U^i\bar{U}^{\bar{j}}U^k\bar{U}^{\bar{l}} \bigg )\nonumber \\
&+& \hf U^i\psi_i+\hf \bar{U}^{\bar{i}}\bar{\psi}_{\bar{i}}
{\bigg \}}\, 
\label{noncq-legendre}
\end{eqnarray}
where the tangent vectors $U^i$ are complex unconstrained superfields, 
and  cotangent vectors
$\psi_i$ are chiral superfields, ${\bar D}_\ad \psi_i=0$.
The variables $U$'s and $\bar U$'s can be eliminated with the aid of their equations 
of motion. This turns the superfield Lagrangian into the hyperk\"ahler potential
\bea
H(\F, \bar \F, \j, \bar \j ) =  K(\Phi,\bar{\Phi})
 &+& \hf \ln \Big(  \L
+ \sqrt{2(\L- g^{\bar{i}\,j}\bar{\psi}_{\bar{i}} \j_j)\,
}
\Big)
-\frac{1}{4} \Big( \L
+ \sqrt{ 2(\L- g^{\bar{i}\,j}\bar{\psi}_{\bar{i}} \j_j)}\,
 \Big)
\non \\
&+& \frac{1}{2} \,\frac{
  \big(g^{\bar{i}\,j}\bar{\psi}_{\bar{i}} \j_j \big)^2
-\frac{1}{4} R^{\bar{i}j\bar{k}l}\bar{\psi}_{\bar{i}}\psi_{j}\bar{\psi}_{\bar{k}}\psi_l 
}
{\L
+ \sqrt{ 2\big(\L- g^{\bar{i}\,j}\bar{\psi}_{\bar{i}} \j_j \big)}\,
}~,
\label{noncq-hcp}
\eea
where 
\bea 
\L = 1 +\sqrt{ 
1 - 2 g^{\bar{i}\,j}\bar{\psi}_{\bar{i}} \j_j 
+2  (g^{\bar{i}\,j}\bar{\psi}_{\bar{i}} \j_j )^2
-\hf R^{\bar{i}j\bar{k}l}\bar{\psi}_{\bar{i}}\psi_{j}\bar{\psi}_{\bar{k}}\psi_l 
}~.
\eea
Here the one-form variables are constrained as follows:
\bea 
1-2 g^{\bar{i}\,j}\bar{\psi}_{\bar{i}}\, \j_j
+ 2(
g^{\bar{i}\,j}\bar{\psi}_{\bar{i}}\, \j_j
)^2 
-\hf R^{\bar{i}j\bar{k}l}\bar{\psi}_{\bar{i}} \,
\psi_{j} \, \bar{\psi}_{\bar{k}} \, \psi_l &>&0~, \non \\
g^{\bar{i}\,j}\bar{\psi}_{\bar{i}}\, \j_j
&< &1~.
\eea
The derivation of the above results can be found in the Appendix
\ref{noncq-hcp-der}.
\\

\noindent
{\bf Acknowledgements:}  
We thank  Muneto Nitta for collaboration at an early 
 stage of this project.
The work of MA  is supported by the bilateral program of Japan Society 
 for the Promotion of Science and Academy of Finland, ``Scientist Exchanges''. 
The work of SMK   is supported  in part
by the Australian Research Council and by a UWA research grant.
The work of UL is supported in part by 
VR grant 621-2003-3454 and by 
EU grant (Superstring theory) MRTN-2004-512194.
SMK is grateful to Dieter L\"ust and Arkady Tseytlin
for hospitality at the University of Munich and Imperial College, 
where some constructions relevant for this work were obtained.

%\vspace{10mm}
\begin{appendix}

\sect{Curvature tensor for  Grassmannians 
and related symmetric spaces}
Using the standard formalism of nonlinear realizations, 
here we compute the curvature 
tensor for the Grassmann manifold and symmetric spaces embedded 
into Grassmannians. Our consideration is  
a streamlined version of Hua's analysis \cite{Hua}.

Using the coset representative (\ref{cosetrep3}), we obtain
\bea
G^{-1} {\rm d}G =E + W~, 
\eea
where $E$ is the vielbein
\bea 
E = \left(
\begin{array}{cc}
0 ~  ~& ~\underline{s} \,{\rm d}\F\,s\\
-s\,{\rm d}\F^\dagger \, \underline{s}  ~~& ~0 
\end{array}
\right)
\equiv 
\left(
\begin{array}{cc}
0 ~  ~& ~\cE\\
-\cE^\dagger  ~~& ~0 
\end{array}
\right)
\label{vielbein}
~,
\eea
and $W$ denotes  the connection
\bea
W = \left(
\begin{array}{cc}
 \underline{s}^{-1} {\rm d}\underline{s}  
 + \underline{s}\, \F{\rm d} \F^\dagger \underline{s}  
~  ~& ~ 0 \\
0  ~~& ~ s^{-1} {\rm d}s +s\,\F^\dagger {\rm d}\F\, s
\end{array}
\right)
~.
\label{connection}
\eea
It is easy to  check that the torsion vanishes,
\bea
T={\rm d}E + W \wedge E + E\wedge W=0~. \label{t-free}
\eea
${}$For the curvature we get 
\bea
R &=&{\rm d} W+ W\wedge W \non \\
&=& \left(
\begin{array}{cc}
 \underline{s}\, {\rm d} \F \wedge s^2 \,{\rm d} \F^\dagger \underline{s}  
~  ~& ~ 0 \\
0  ~~& ~ s \,{\rm d}\F^\dagger \wedge \underline{s}^2  \,
{\rm d}\F\, s
\end{array}
\right) 
=
 \left(
\begin{array}{cc}
\cE\wedge \cE^\dagger
~  ~& ~ 0 \\
0  ~~& ~ 
\cE^\dagger \wedge \cE
\end{array}
\right) ~.
\label{curvature-comp}
\eea

It is quite transparent that the above results 
apply to the Hermitian symmetric spaces  
${\rm SO}(2n)/{\rm U}(n)$ and ${\rm Sp}(n)/{\rm U}(n)$
simply by restricting $\F$ to obey the appropriate 
algebraic symmetry conditions.

Similar calculations can be performed in the  non-compact case. 
${}$For the symmetric space ${\rm U}(n,m)/{\rm U}(n)\times {\rm U}(m)$, 
the coset representative is given by eq. (\ref{nc-cosetrep3}). 
One obtains 
\begin{eqnarray}
 G^{-1} {\rm d}G=E+W~,
\end{eqnarray} 
where the vielbein is 
\begin{eqnarray}
 E=\left(
\begin{array}{cc}
 0 & \underline{s}\,{\rm d} \F \,s\\
 s \,{\rm d} \F^\dagger \underline{s} & 0
\end{array}
\right)
\equiv \left(
\begin{array}{cc}
0 ~  ~& ~\cE\\
\cE^\dagger  ~~& ~0 
\end{array}
\right)~,~~~~
\label{vielbein-non}
\end{eqnarray}
and the connection has the form
\begin{eqnarray}
 W=\left(
\begin{array}{cc}
 \underline{s}^{-1}{\rm d}\underline{s}
-\underline{s}\, \F{\rm d}\F^\dagger\underline{s} ~~& 0\\
 0 &  s^{-1}{\rm d}s
-s \,\F^\dagger {\rm d}\F s 
\end{array}
\right)~.
\end{eqnarray}
The corresponding geometry is again torsion-free,
\begin{eqnarray}
 T={\rm d}E + W \wedge E + E\wedge W=0~.
\label{t-free-non}
\end{eqnarray}
The curvature can be shown to be
\begin{eqnarray}
R&=&{\rm d} W+ W\wedge W \nonumber \\
&=&\left(
\begin{array}{cc}
-\underline{s}\,{\rm d}\F s \wedge s {\rm d}\F^\dagger \underline{s} ~~&  0\\
0 & -s\,{\rm d}\F^\dagger \underline{s}
\wedge \underline{s} {\rm d}\F s
\end{array}
\right)
=\left(
\begin{array}{cc}
-\cE\wedge \cE^\dagger
~  ~& ~ 0 \\
0  ~~& ~ 
-\cE^\dagger \wedge \cE
\end{array}
\right)~.
\label{curvature-non-comp}
\end{eqnarray}

The results obtained for  ${\rm U}(n,m)/{\rm U}(n)\times {\rm U}(m)$
remain valid for the Hermitian symmetric spaces  
${\rm SO}^*(2n)/{\rm U}(n)$ and 
 ${\rm Sp}(n,{\mathbb R})/{\rm U}(n)$
if one restricts $\F$ to obey the appropriate 
algebraic symmetry conditions.

\sect{Derivation of (\ref{noncq-hcp})}
\label{noncq-hcp-der}
This appendix is devoted to the derivation of the hyperk\"ahler potential 
(\ref{noncq-hcp}). 
Since the base manifold is a symmetric space,
it is sufficient to implement the dualization, 
for the action (\ref{noncq-legendre}),
at $\F=0$.
The first-order Lagrangian 
\bea
\cL =-\hf \ln \O + \hf U^i \J_i +\hf {\bar U}^i \J_i~, \qquad 
\O = 1 +2 U^\dagger U + |U^{\rm T}U|^2 
\eea 
leads to the following equations of motion for  ${\bar U}$'s and $U$'s:
\bea
\frac{ U^i +{\bar U}^i U^{\rm T}U }{\O} =\hf {\bar \J}_i~, \qquad 
\frac{ {\bar U}^i + U^i \overline{U^{\rm T}U }}{\O} =\hf \J_i~,
\eea
where $\J$ is a cotangent vector at $\F=0$.
These equations imply 
\bea
\frac{1}{4} \J^{\rm T}\J =\frac{ \overline{U^{\rm T}U }}{\O}~, 
\qquad 
\frac{1}{4} \overline{\J^{\rm T}\J }=\frac{{U^{\rm T}U }}{\O}~,
\eea
and also 
\bea
\frac{1}{16} \Big( 1-2 \J^\dagger \J+ |\J^{\rm T} \J|^2 \Big)
= \Big( \frac{U^\dagger U}{\O} -\frac{1}{4}\Big)^2~.
\label{UdaggerU}
\eea
The latter is consistent if 
\bea
 1-2 \J^\dagger \J+ |\J^{\rm T} \J|^2 >0~.
 \eea
 By construction, the correspondence between 
the tangent and cotangent variables 
 should be such that $\U \to 0 \Longleftrightarrow \J \to 0$. 
This means that we are to choose
 the ``minus'' solution of (\ref{UdaggerU}), that is 
\bea
 \frac{U^\dagger U}{\O} =\frac{1}{4}\Big( 1 -
 \sqrt{ 1-2 \J^\dagger \J+ |\J^{\rm T} \J|^2 } \Big)~.
 \eea

Now, the results obtained above can be used to express $\O$ via $\J$
and its conjugate. By definition, we have 
\bea
\frac{1}{\O} = \frac{1}{\O^2} +\frac{2}{\O} \, \frac{U^\dagger U}{\O} 
+\Big|\frac{{U^{\rm T}U }}{\O}\Big|^2~.
\eea
This is equivalent to 
\bea
\Big( \frac{1}{\O} 
-\frac{1}{4}
\L
\Big)^2
=\frac{1}{16} \Big(
\L^2- |\J^{\rm T} \J|^2 \Big)~,
\label{1overL}
\eea 
where
\bea
\L = 1 +  \sqrt{ 1-2 \J^\dagger \J+ |\J^{\rm T} \J|^2 } ~.
\eea
The consistency of eq. (\ref{1overL}) can be seen to require
\bea
\J^\dagger \J <1~.
\eea
Since for $\J \to 0$ we should have $\O \to 1$, 
it is necessary to choose the ``plus'' solution of (\ref{1overL}), that is
\bea
\frac{4}{\O} = \L + \sqrt{ \L^2 - |\J^{\rm T} \J|^2 }
=\L +\sqrt{2(\L -\J^\dagger \J)}
~.
\eea

The above consideration corresponds to the origin, 
$\F=0$, of the base manifold. To extend these results 
to an arbitrary point $\F$ of the base manifold, we should replace 
\bea
 \J^\dagger \J ~\to~ g^{\bar{i}\,j}\bar{\psi}_{\bar{i}} \j_j ~,
\qquad 
 |\J^{\rm T} \J|^2 ~ \to ~ 
 2(g^{\bar{i}\,j}\bar{\psi}_{\bar{i}}\, \j_j)^2 
-\hf R^{\bar{i}j\bar{k}l}\bar{\psi}_{\bar{i}} \,
\psi_{j} \, \bar{\psi}_{\bar{k}} \, \psi_l ~.
\eea

\end{appendix}

%%%%%%%%%%%%%%%%%%%%%%%%%%%%%%%%%%%%%%%%%%%%%%%%%%%%%
%
% references
%
%%%%%%%%%%%%%%%%%%%%%%%%%%%%%%%%%%%%%%%%%%%%%%%%%%%%%

%%
\small{

}
\end{document}